\documentclass{jfm}

\usepackage{graphicx}
\usepackage{epstopdf,epsfig}
\usepackage{newtxtext}
\usepackage{subfigure}
\usepackage{newtxmath}
\usepackage{longtable}
\usepackage{xcolor}
\usepackage{booktabs}
\usepackage{natbib}
\usepackage{hyperref}
\hypersetup{
    colorlinks = true,
    allcolors=blue,
}

\newcommand{\RomanNumeralCaps}[1]
\nnlinenumbers

\graphicspath{{Figs/}}

\title{Scaling behaviour of rotating convection in a spherical shell with different Prandtl numbers}

\author{Wei Fan\aff{1}, Qi Wang\aff{1}
   and
  Yufeng Lin \aff{1,2}
 \corresp{\email{linyf@sustech.edu.cn}}}

\affiliation{\aff{1} Department of Earth and Space Sciences, Southern University of Science and Technology, Shenzhen, 518055, PR China
\aff{2} Center for Complex Flows and Soft Matter Research, Southern University of Science and Technology, Shenzhen, 518055, PR China}

\begin{document}
\maketitle

\begin{abstract}
Rayleigh-Bénard convection in a rotating spherical shell provides a simplified model for convective dynamics of planetary and stellar interiors. Over the past decades, the problem has been studied extensively via numerical simulations, but most of previous simulations set the Prandtl number $\Pran$ of unity.
In this study, we build more than 200 numerical models of rotating convection in a spherical shell over a wide range of $\Pran$ ($10^{-2}\le\Pran\le10^2$). As increasing the Rayleigh number $Ra$, we characterise four different flow regimes, starting from the linear onset to multiple modes, then transiting to the geostrophic turbulence and eventually approaching the weakly rotating regime. In the multiple modes regime, we show evidence of triadic resonances in numerical models with different $\Pran$, which may provide a generic mechanism for the transition from laminar to turbulence in rotating convection.  
We analyse scaling behaviours of the heat transfer and convective flow speeds in numerical simulations, paying particular attention to the $\Pran$-dependence. We find that the so-called diffusion-free scaling for the heat transfer cannot reconcile all numerical models with different $\Pran$ in the geostrophic turbulence regime. However, the characteristic flow speeds at different $\Pran$ roughly follow a unified scaling that can be described by VAC (Visco-Archimedean-Coriolos) force balances, though the scaling tends to approach the CIA (Coriolis–Inertial–Archimedean) force balance at low $\Pran$. We also show that transition behaviours from rotating to non-rotating convection depend on $\Pran$. The transition criteria based on heat transfer and flow morphology would be rather different when $\Pran>1$, but the two criteria are consistent for cases with $\Pran\le 1$. Both scaling behaviours and transition behaviours suggest that the heat transfer is controlled by the boundary layers while the convective flow speeds are mainly determined by the force balance in the bulk for cases with $\Pran>1$, which is in line with recent experimental results with moderate to high $\Pran$. For cases with $\Pran \le 1$, both the heat transfer and convective velocities are approaching the inviscid dynamics in the bulk. We also briefly analyzed the magnitude and scaling of zonal flows at different $\Pran$, showing that the zonal flow amplitude rapidly increases as $\Pran$ decreases.
\end{abstract}

\section{Introduction}
\label{sec:Introduction}
 Rotation plays an important role in many natural convection systems such as Earth's liquid outer core, convective envelopes of giant planets and rotating stars \citep{aurnou2015rotating,jones2015thermal}. Rotating Rayleigh-Bénard convection (RRBC) provides a canonical model for studying convective dynamics under the influence of rotation \citep{Chandra1961}. The RRBC problem
 has been extensively studied for many decades using theoretical analysis, numerical simulations and laboratory experiments \citep[e.g. see review articles][]{ecke2023turbulent,Xia2023}. While the RRBC model captures some aspects of the key dynamics of rotating convection in a local region, it is  important to consider the effect of spherical geometry for the global dynamics of planetary and stellar interiors.  
 
 Theoretical works on spherical rotating convection concentrated on the onset of convection \citep{Chandra1961,Roberts1968,Busse1970,zhang1994coupling,Jones2000,dormy2004onset}. A few experiments were conducted to model spherical rotating convection using centrifugal gravity \citep[e.g.][]{Busse1976,Cardin1994,aubert2001systematic}, but it is difficult to explore a wide range of parameter regime due to practical considerations. The spherical rotating convection in the highly supercritical regime has been mainly studied using direct numerical simulations \citep[e.g.][]{gilman1977nonlinear,tilgner1997finite, christensen2002zonal,gillet2006quasi,Aurnou2007,yadav2016effect,Guervilly2019}. Several scaling laws of rotating convection have been proposed based on heuristic arguments \citep{stevenson1979turbulent,Aurnou2020} and tested using numerical models in spherical geometries \citep{gastine2016scaling, long2020scaling,Lin2021}. However,  
 most of previous simulations set the $\Pran$ of unity, which may mask some scaling behaviours depending on $\Pran$. It is important to consider the $\Pran$-dependence when extrapolating scaling laws to planetary cores because thermal and compositional diffusivities are different in planetary core conditions \citep{labrosse2003thermal,li2000chemical,zhang2020reconciliation}, corresponding to quite different $\Pran$  for thermal convection ($\Pran\sim 0.01$) and compositional convection ($\Pran\sim 10$).   
 In this study, we perform a set of numerical simulations of rotating convection in a spherical shell over a wide range of $\Pran$ ($10^{-2}\le\Pran\le10^2$).       

There are two branches of rotating convection near the onset depending on $Pr$, namely viscous convection and inertial convection \citep{zhang2017theory}.  
 At large $\Pran$ ($\Pran\ge 1$), the onset of convection invokes the viscous force and  is in the form of quasi-steady columnar rolls \citep{Busse1970,Jones2000}.
 The critical Rayleigh number for the viscous convection scales as $Ra_c \sim E^{-4/3}$, and the azimuthal wavenumber of the onset mode $m_c\sim E^{-1/3}$, where $E$ is the Ekman number. At low $\Pran$ ($\Pran \ll 1$), the onset of convection invokes the inertial term and is in the form of oscillatory inertial modes \citep{zhang1994coupling}. The critical Rayleigh number for the inertial convection scales as $Ra_c \sim E^{-1/2}$ and the azimuthal wavenumber of the onset mode $m_c\sim O(1)$ \citep{zhang2017theory}. Of course, there always exists a transition from viscous convection to inertial convection depending on $E$ and $\Pran$, where the onset is characterised by spiralling columnar convection \citep{Zhang1992}. 
 More recently, it has been found that the subcritical convection can occur at low $\Pran$ \citep{Guervilly2016,Kaplan2017}. These findings highlighted the significant role of $\Pran$ in rotating convection near the onset. 
 
 In the highly supercritical regime, the rigorous theoretical analysis is not tractable. Nevertheless, some scaling laws regarding the heat transfer efficiency, the typical convective flow speed and length scale have been proposed relying on heuristic arguments and force balances \citep{stevenson1979turbulent, aubert2001systematic,gillet2006quasi,barker2014theory, Aurnou2020}. For rotating turbulent convection that is the most relevant regime to the planetary core dynamics, the so-called diffusion-free scalings or inertial scalings have been widely used to compare experimental and numerical results \citep{Cheng2018,hawkins2023laboratory}.
 The diffusion-free scalings were derived using various approaches in the literature, but these scalings are essentially based on the Coriolis–Inertial–Archimedean (CIA) force balance and incorporated with the mixing length theory, leading to scalings independent of the fluid viscosity and thermal diffusivity \citep{stevenson1979turbulent,julien2012heat,jones2015thermal}. 
 The diffusion-free scalings predict the heat transfer efficiency measured by the Nusselt number $Nu\sim Ra^{3/2}E^2 \Pran^{-1/2}$, and the convective flow speed measured by the non-zonal Reynolds number $Re_{non} \sim (Ra_{Q}\Pran^{-2}E^{1/2})^{2/5}$, where $Ra_Q$ is the flux-based Rayleigh number $Ra_{Q}=(Nu-1)Ra$. These scalings exhibit $\Pran$-dependence, but most previous numerical studies mainly considered the dependence on $Ra$ and $E$. These scalings were tested using direct numerical simulations over wide range of $Ra$ and $E$ in rotating spherical shells with the fixed temperature \citep{gastine2016scaling}  and  fixed-flux \citep{long2020scaling}  boundary conditions, and in a uniformly heated full sphere \citep{Lin2021}. More recently, \citet{wang2021diffusion,gastine2023latitudinal} examined the latitudinal dependence of the heat transfer in rotating spherical shells. \citet{gastine2023latitudinal} found that the heat transfer scaling in the polar region is steeper than the diffusion-free scaling, whereas the scaling in the equatorial region is less steep than the diffusion-free scaling. \cite{wang2021diffusion} showed that the diffusion-free scalings may be valid in the mid-latitude flow region. Note that some recent studies in cylindrical domains also reveal that the heat transfer scaling has a radius dependence due to difference between bulk convective flow and sidewall dynamics \citep{zhang2020boundary,zhang2021boundary,de2020turbulent}. Most of these numerical simulations adopt $\Pran=1$, which corresponds to a special parameter space as we shall show in this study. 
 
 Previous studies on the $\Pran$ effects in Rayleigh-Bénard convection have shown different scaling behaviours depending on $Pr$. \citet{verzicco1999prandtl}  investigated $Nu$ as a function of $\Pran$ within the interval of $[0.0022, 15]$ using direct numerical simulations and found that $Nu \sim \Pran^{0.14}$ for $\Pran \le 0.35$, whereas $Nu$ is almost independent of $Pr$ for $\Pran>0.35$. Other studies in different $\Pran$ intervals shown similar behaviors \citep[e.g.][]{kerr2000prandtl,silano2010numerical,Li2021}. In RRBC, \cite{Zhong2009} found the heat transfer behaviour for small $Pr\le 0.7$ differs from moderate $Pr$. \cite{King2013} also found that the transition from rotating to non-rotating convection of liquid metal substantially differs from that of water. More recently, \cite{Abbate2023} carried out a suite of RRBC experiments in moderate to high $\Pran$ fluids, which showed that the bulk interior flows can be described by the CIA force balance, but the heat transfer is controlled by the boundary layers, the similar behaviors was also found in plane layer \citep{oliver2023small}. The effects of $Pr$ in spherical rotating convection have been considered experimentally \citep{aubert2001systematic} and numerically \citep{tilgner1997finite,christensen2002zonal,gillet2006quasi}. They found that small $\Pran<1$ tends to reduce the heat transfer and promote zonal flows. In particular, \citet{gillet2006quasi} noticed that the diffusion-free scaling does not represent the $\Pran$-dependence. These studies pointed out that the value of $\Pran$ plays an important role in spherical rotating convection not only in the onset regime but also in the highly supercritical regime.
 
 In this study, we build more than 200 numerical models over a wide range of $\Pran$ ($10^{-2}\le\Pran\le10^2$) to investigate the influence of $\Pran$ on the rotating convection in a spherical shell. We show that the diffusion-free scaling for the heat transfer roughly fits the numerical results with $\Pran\le 1$ but has problems to reconcile  numerical models with $\Pran > 1$ in the geostrophic turbulent regime. On the other hand, the convective flow speeds with different $\Pran$ in the geostrophic turbulence regime roughly follow a unified scaling. We also show that the transition behaviours from rotating to non-rotating convection depend on $\Pran$ \citep{King2013}. We find that the transition criteria based on heat transfer and flow morphology are different for cases with $\Pran>1$, in which the heat transfer starts to deviate from the rotating scaling as increasing $Ra$, but the flow structures remain geostrophic. These findings are in line with recent laboratory experiments at moderate to high $\Pran$, which suggested that the interior flows are controlled by the force balance in the bulk but the heat transfer is controlled by the boundary layers  \citep{Abbate2023}. On the other hand, our numerical models at low $\Pran$ suggest that both heat transfer and convective flow speeds are controlled by the dynamics in the bulk.

\section{Numerical models}\label{sec:Numerical models}

\subsection{Governing equations}
We consider Boussinesq convection of homogeneous fluid in a spherical shell of inner radius $r_i$ and outer radius $r_o$ that uniformly rotates at $\boldsymbol \Omega=\Omega \boldsymbol{\hat z}$. Convection is driven by fixed temperature difference $\Delta T=T_{i}-T_{o}$ between the inner and outer boundaries, under the gravity $\boldsymbol{g}=-g_o \boldsymbol{r}/r_o$. Using the shell thickness $D = r_{o} - r_{i}$ as the length scale,  $\Omega^{-1}$ as the time scale, $\Delta T$ as the temperature scale and the gravity at the outer boundary $g_o$ as the reference value, the dimensionless governing equations can be expressed as
\begin{equation}
  \frac{\partial \boldsymbol u}{\partial t}+ \boldsymbol u \cdot \boldsymbol \nabla\boldsymbol u+2 \hat{\boldsymbol z} \times \boldsymbol u = -\boldsymbol \nabla P+E \boldsymbol \nabla^{2} \boldsymbol u+ \frac {RaE^2}{Pr} T \boldsymbol r,
  \label{eq21}
\end{equation}

\begin{equation}
  \frac{\partial T}{\partial t}+\boldsymbol u \cdot \boldsymbol \nabla T =\frac{E}{\Pran} \boldsymbol \nabla^{2} T,
  \label{eq22}
\end{equation}
\begin{equation}
  \boldsymbol \nabla \cdot \boldsymbol u=0,
  \label{eq23}
\end{equation}
where $\boldsymbol{u}$ is the velocity, $P$ is the reduced pressure and $T$ is the temperature. The system is defined by three dimensionless control parameters, the Ekman number $E$, the Rayleigh number $Ra$ and the Prandtl number $\Pran$: 
\begin{equation}
  E=\frac{\nu}{\Omega D^2},\quad Ra=\frac{\alpha g_{0}\Delta T D^3}{\nu \kappa}, \quad  \Pran=\frac{\nu}{\kappa},
  \label{eq24}
\end{equation}
 where $\nu$ is the kinematic viscosity, $\kappa$ is the thermal diffusivity and $\alpha$ is the thermal expansion coefficient. The rotational modified Rayleigh number $Ra^*$ is often used in the rotating convection and $Ra^*$ is also the squared convective Rossby number, which is used in many rotating convection studies \citep[e.g.][]{Gilman1977,gastine2013solar}.
\begin{equation}
  Ra^*=\frac{\alpha_{T} g_{0}\Delta T}{\Omega^2 D}=\frac{RaE^2}{\Pran}={Ro_c}^2,
  \label{eq25}
\end{equation}
which provides a clear relationship between thermal buoyancy and Coriolis force in rotating convection.

In this study, we fix the radius ratio of the shell $\eta=r_i/r_o=0.35$. Both boundaries are impermeable, no-slip and held at constant temperatures.

 \subsection{Numerical technique}
 We use the open-source code XSHELLS \href{https://www.bitbucket.org/nschaeff/xshells/}{( https://www.bitbucket.org/nschaeff/xshells/)} to solve the governing equations (\ref{eq21}-\ref{eq23}) subjected to the boundary conditions. The fluid is assumed to be incompressible, and the velocity $\boldsymbol u$ can be decomposed into toroidal and poloidal components:
\begin{equation}
\boldsymbol u=\boldsymbol \nabla \times(\mathcal{T} \boldsymbol r)+\boldsymbol \nabla \times \boldsymbol \nabla \times(\mathcal{P} \boldsymbol r).
\label{eq26}
\end{equation}
The toroidal $\mathcal{T}$, poloidal $\mathcal{P}$ scalar fields, and the temperature field $T$ are expanded in terms of spherical harmonic expansion on spherical surfaces. XSHELLS uses second-order finite differences method in radial direction and pseudo-spectral spherical harmonic expansion. The spectral expansion is truncated up to spherical harmonics of degree $L_{max}$ and $Nr$ denotes the number of radial gird points. The time-stepping scheme is second order, and treats the diffusive terms implicitly, while the nonlinear and Coriolis terms are handled
explicitly. This code also uses the SHTns library  to speed up the spherical harmonic transformations \citep{schaeffer2013efficient}.

\subsection{Diagnostics}\label{sec:diagnostics}
We analyze several diagnostics properties to quantify the influence of different control parameters on heat and momentum transports. We adopt several notations regarding averaging procedures. Overbars ${\overline{\dots}}$ correspond to temporal averaging, angular brackets $\langle \dots\rangle$ to spatial averaging over the entire spherical shell volume and $\langle \dots\rangle_s$ to an average over a spherical surface:
\begin{equation}
  \overline{f}=\frac{1}{\tau} \int_{t_{0}}^{t_{0}+\tau}f \mathrm{d}t, \langle f\rangle=\frac{1}{V} \int_{V} f(r, \theta, \phi) \mathrm{d} V, \langle f\rangle_s=\frac{1}{4 \pi} \int_{0}^{2 \pi} \int_{0}^{\pi} f(r, \theta, \phi) \sin \theta \mathrm{d} \theta \mathrm{d} \phi.
  \label{eq27}
\end{equation}
where $\tau$ is the time averaging interval, $V$ is the volume of the spherical shell, $r$ is the radius, $\theta$ is the colatitude and $\phi$ is the longitude. 

The Nusselt number $ Nu$ denotes heat transport, the ratio of the total heat flux to the conduction heat flux. In spherical shells, the conductive temperature profile $T_{c}$ is the solution of
\begin{equation}
 \frac{\mathrm{d} T_{c}}{\mathrm{~d} r} = -\frac{r_{i}r_{o}}{r^2}, \quad T_{c}\left(r_{i}\right) = 1, \quad T_{c}\left(r_{o}\right) = 0.
 \label{eq28}
\end{equation}
Following \citet{gastine2015turbulent}, it yields
\begin{equation}
 T_{c}(r)=\frac{\eta}{(1-\eta)^2}\frac{1}{r}-\frac{\eta}{1-\eta},
 \label{eq29}
\end{equation}
where $\eta=r_{i}/r_{o}$ is the radius ratio. The notation $\vartheta$ is introduced to define the time and horizontally averaged radial dimensionless temperature profile
\begin{equation}
\vartheta(r)=\overline{\langle T\rangle_s}.
\label{eq30}
\end{equation}
Then the Nusselt number is
\begin{equation}
Nu=-\eta\frac{\mathrm{d}\vartheta}{\mathrm{d}r}(r=r_{i})=-\frac{1}{\eta}\frac{\mathrm{d}\vartheta}{\mathrm{d}r}(r=r_{o}).
\label{eq31}
\end{equation}
Here we only consider the global averaged $Nu$, but the efficiency of convective heat transfer slightly varies as latitude in rotating convection \citep{wang2021diffusion,gastine2023latitudinal}. The total kinetic energy is given by
\begin{equation}
  E_{k}=\frac{1}{2} \int_{v} u^2 \mathrm{d}V=\sum_{l = 1}^{l_{\max }} \sum_{m = 0}^{l} \mathcal{E}_{l}^{m}(t),
  \label{eq32}
\end{equation}
where $\mathcal{E}_{l}^{m}$ is the dimensionless kinetic energy at a spherical harmonic degree $l$ and order $m$.The total kinetic energy can be decomposed into zonal and non-zonal parts:
\begin{equation}
  E_{zon}=\frac{1}{2} \int_{v}{(u_{\phi}^{0})}^2 \mathrm{d}V, \quad E_{non}=E_{k}-E_{zon}.
  \label{eq33}
\end{equation}
We define the Rossby number as
\begin{equation}
  Ro=\frac{U_{rms}}{\Omega D},
  \label{eq34}
\end{equation}
where $U_{rms}$ is the dimensional root-mean-square (r.m.s.) velocity. Based on the non-dimensionalisation we used, the Rossby number can be determined through kinetic energy as
\begin{equation}
  Ro=\sqrt{\frac{2E_{k}}{V}},
  \label{eq35}
\end{equation}
Accordingly, we have the non-zonal Rossby number
\begin{equation}
  Ro_{non}=\sqrt{\frac{2E_{non}}{V}}.
  \label{eq36}
\end{equation}
and the zonal Rossby number is
\begin{equation}
  Ro_{zon}=\sqrt{\frac{2E_{zon}}{V}}.
  \label{eqrozon}
\end{equation}
We also define the local Rossby number $Ro_{\ell}$
\begin{equation}
  Ro_{\ell}=\frac{U_{rms}}{\Omega \ell}=Ro\frac{D}{\ell},
  \label{eq37}
\end{equation}
where $\ell$ is the typical flow length scale and is determined from the time-averaged kinetic energy spectrum following \citet{christensen2006scaling}
\begin{equation}
\ell^{-1} = \overline{\left(\frac{D \sum\limits_{l = 1}^{l_{\max }} \sum\limits_{m = 0}^{l} l \mathcal{E}_{l}^{m}(t)}{\pi \sum\limits_{l = 1}^{l_{\max }} \sum\limits_{m = 0}^{l} \mathcal{E}_{l}^{m}(t)}\right)}.
\label{eq38}
\end{equation}
In rotating spherical shell convection, the zonal flow component does not contribute to the heat transport from the inner to the outer shell, so we focus on the non-zonal component and it can be determined through non-zonal Rossby number as
\begin{equation}
  Re_{non}=\frac{Ro_{non}}{E}.
  \label{eq39}
\end{equation}
and the zonal Reynolds number is
\begin{equation}
  Re_{zon}=\frac{Ro_{zon}}{E}.
  \label{eqrezon}
\end{equation}

\section{Numerical results}\label{sec:Numerical results}
\subsection{Overview}\label{subsec:OV}
 \begin{figure}
  \centerline{\includegraphics[width=0.95 \textwidth]{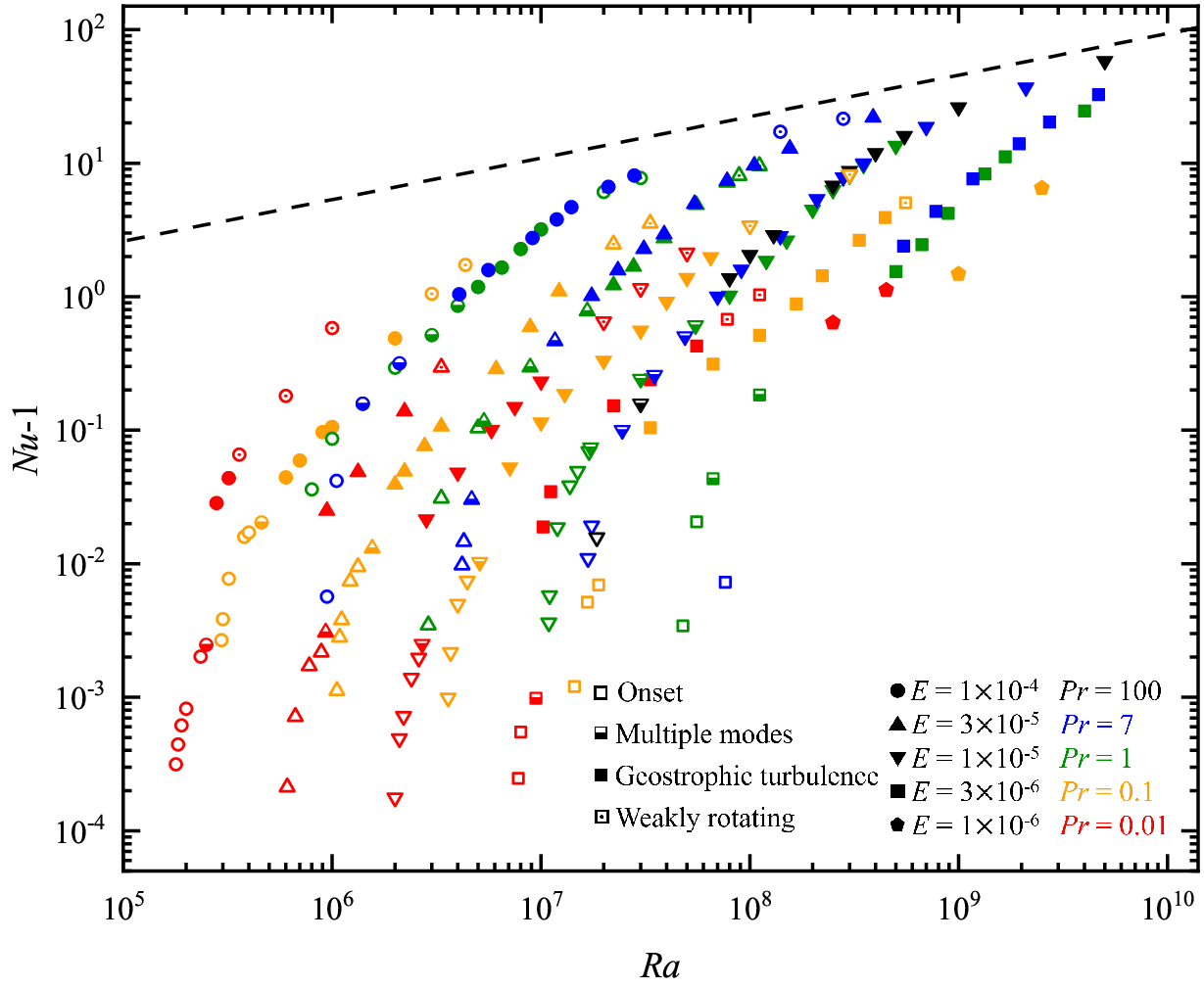}}
  \caption{Regime diagram of all of numerical models in the parameter space $(Nu-1, Ra)$. Different colors and shapes correspond to different $\Pran$ and $E$ respectively.  Open symbols denote the onset (ON) regime; half-filled symbols denote the multiple modes (MM) regime; filled symbols denote the geostrophic turbulent (GT) regime;  central dot symbols denote the weakly rotating (WR) regime.  The dashed line corresponds to the non-rotating scaling $(Nu-1)\propto Ra^{1/3}$.}
\label{fig:all}
\end{figure}

\begin{figure}
  \centerline{\includegraphics[width=13.1cm,height=10cm]{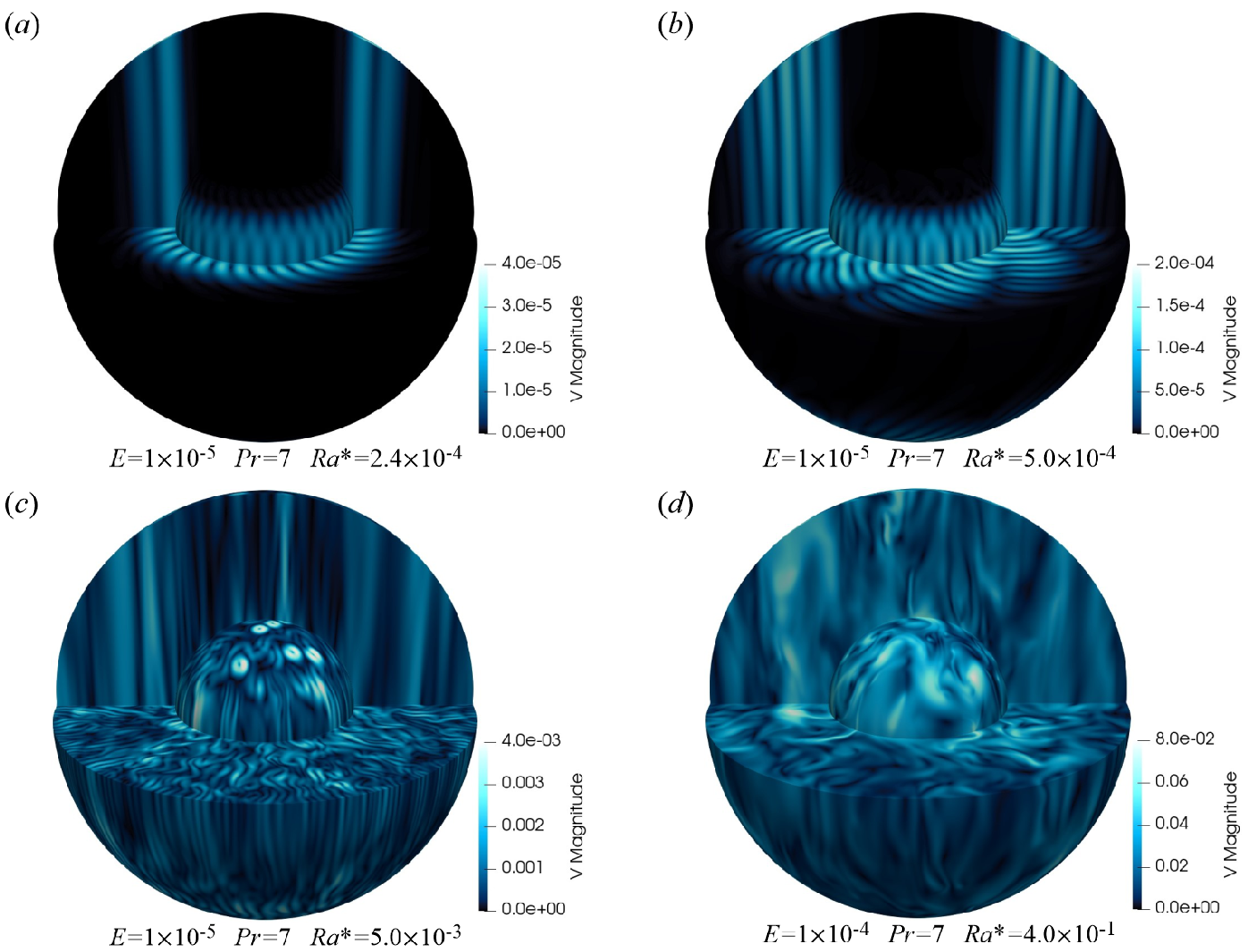}}
  \caption{ Contours of velocity magnitude in units of the Rossby number at meridional and equatorial planes, and spherical surfaces showing typical flow structures of four different regimes with $\Pran=7$. The inner (outer) surface corresponds to spherical surface of radius $r_i+0.1D$ ($r_o-0.1D$). The color bar represents Rossby number $Ro$. $(a)$ onset (ON) regime; $(b)$ multiple modes (MM) regime; $(c)$ geostrophic turbulent (GT) regime; $(d)$ weakly rotating (WR) regime.}
\label{fig:3dlargePr}
\end{figure}

\begin{figure}
  \centerline{\includegraphics[width=13cm,height=10cm]{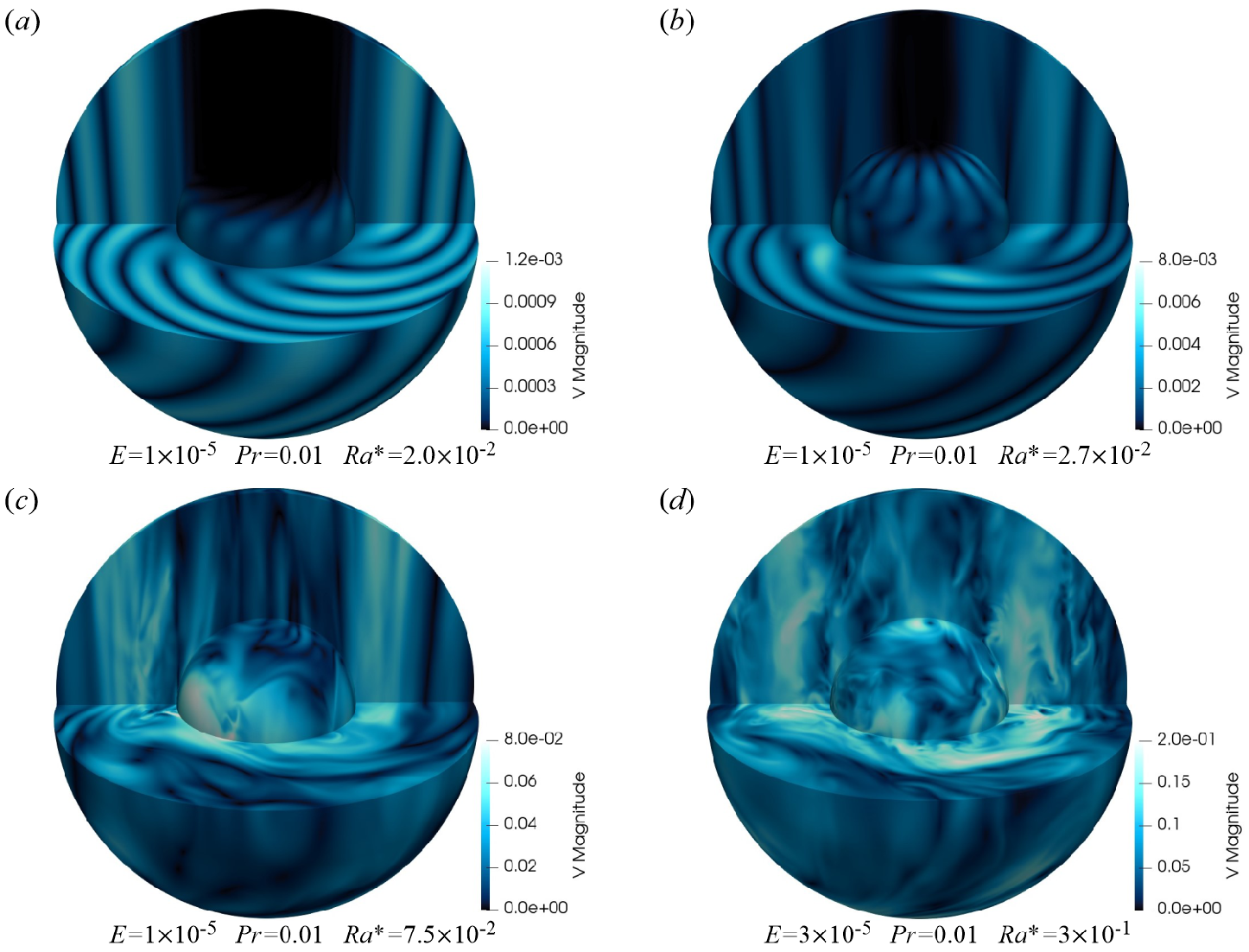}}
  \caption{Same as figure \ref{fig:3dlargePr} but for cases with $\Pran=0.01$. $(a)$ onset (ON) regime; $(b)$ multiple modes (MM) regime; $(c)$ geostrophic turbulent (GT) regime; $(d)$ weakly rotating (WR) regime.}
\label{fig:3dsmallPr}
\end{figure} 

We have simulated a total of 211 cases in the parameter space of $1.79\times 10^{5} \le Ra \le 5\times 10^{9}$, $1\times10^{-6} \le E \le 1\times10^{-4}$ and $0.01 \le \Pran \le 100$.  Details of input and diagnostic parameters for all models are listed in table \ref{tab:data} in the Appendix \ref{AppA}. Figure \ref{fig:all} shows $Nu-1$ as a function of $Ra$ for all the numerical models. We use different colors and shapes to distinguish different values of $\Pran$ and $E$ respectively. Broadly speaking, the ($Nu-1, Ra$) diagram shows that $Nu$ is insensitive to $Pr$ when $\Pran \ge 1$ (black, blue and green symbols) but highly depends on $Pr$ when $\Pran < 1$ (orange and red symbols). This is in line with previous studies on the $\Pran$ effects in Rayleigh-B\'enard convection \citep{verzicco1999prandtl}, and suggests that we need to consider different scaling behaviours for small $\Pran$ and large $\Pran$.    

We also see from figure \ref{fig:all} that slopes of $Nu-1$ change as increasing $Ra$ because of different convective regimes. Our numerical models can be separated into four different regimes depending on the flow morphology and heat transport efficiency. When the Rayleigh number is slightly above the critical value $Ra_c$, convection is characterised by a single onset mode, which is referred to as the onset (ON) regime (open symbols in figure \ref{fig:all}). The critical $Ra_c$ and the structure of onset mode depend on $E$ and $Pr$ (see table \ref{tab:critical} in Appendix \ref{AppB}). In the onset regime, weak convection has little contribution to the heat transfer, so we can see that $(Nu-1)\ll 1$ but increases steeply as a function of $Ra$. As we increase $Ra$, the convective flow remains laminar, but several convective modes can coexist, which is referred to as the multiple modes (MM) regime (half-filled symbols in figure \ref{fig:all}).
In the MM regime, the Nusselt number remains small, i.e. $(Nu-1)<1$, meaning that heat is mainly transferred by conduction. 
Further increasing $Ra$, convection becomes more complex and even turbulent but exhibits columnar structures along the rotation axis, which is referred to as the geostrophic turbulence (GT) regime (filled symbols in figure \ref{fig:all}). As the transition from laminar to turbulent flow  is not always well defined, it is not straightforward to define the boundary between the MM and GT regimes. There exists sudden jumps of $Nu-1$ as increasing $Ra$ for $\Pran<1$ cases (see red and orange symbols in figure \ref{fig:all}), so we define such jumps as the boundary between the MM and GT regimes when $\Pran<1$. However, $Nu-1$ smoothly increases as increasing $Ra$ when $Pr\ge 1$, so we simply define $Nu>2$ as the criterion to enter the GT regime for $Pr\ge 1$ cases following \cite{gastine2016scaling}. This criterion amounts to say that the heat transported by convection overtakes the conductive heat transfer. When $Ra$ is sufficiently large for fixed $E$ and $\Pran$, convection becomes less geostrophic by breaking the rotational constraint and eventually approaches non-rotating convection. Again it is not straightforward to define the transition from rotating to non-rotating turbulence. Some scalings and criteria have been proposed to characterise the transition from rotating to non-rotating convection \citep{julien2012heat,gastine2016scaling,long2020scaling}, but it is difficult to reconcile numerical simulations with different $Pr$ as we shall show. Here we simply use the local Rossby number $Ro_\ell>0.1$ as the criterion for the weakly rotating (WR) regime (central dot symbols in figure \ref{fig:all}). The local Rossby number is the ratio of advection to Coriolis force and  $Ro_\ell \lesssim 0.1$ is often seen as an indicator of rotating flows \citep[e.g.][]{Davidson2014}. This criterion is a bit subjective, but we do see the columnar structures tend to break when $Ro_\ell > 0.1$. We define different flow regimes to facilitate our following discussions, but determining the regime boundaries is not the focus of this study.

Figures \ref{fig:3dlargePr} and \ref{fig:3dsmallPr} show typical flow structures of four different regimes for cases with $\Pran=7$ and $\Pran=0.01$ respectively. We have mentioned that the heat transfer exhibit different behaviors between models with $\Pran\ge 1$ and $\Pran< 1$. Such difference is also reflected in the flow structures in all regimes. In the onset regime, the convective mode at $\Pran=7$ takes the form of columnar rolls in the vicinity of the tangent cylinder \citep{dormy2004onset,Barik2023}, while the convection at $\Pran=0.01$ shows spiralling columnar structures that occur in the whole domain outside the tangent cylinder \citep{Zhang1992}. The spiralling structure corresponds to the transitional onset between the viscous convection and inertial convection\citep{zhang2017theory}. In the MM regime, the convective flows are similar to the onset modes but apparently show other unstable modes coexisting. We will show in \S\ref{sec:MM} that the interaction of multiple modes may take place in triadic resonances \citep{lin2021triadic}. In the GT regime, convection flows are chaotic in the equatorial plane but are organized along the rotation axis. The horizontal length scale is dramatically different between cases with $\Pran=7$ and $\Pran=0.01$ in the same $E$. In the case of $E=1\times 10^{-5}$ and $\Pran=7$, the range of $\ell$ is $0.064 -0.089$. However, when $\Pran=0.01$, the range of $\ell$ is $0.488 - 0.529$. In the WR regime, columnar convective structures are broken as the rotational constraint becomes less important.
Figures \ref{fig:3dlargePr} and \ref{fig:3dsmallPr} provide an overview of the flow morphology in different regimes. In the following, we will focus on the scaling behaviors of the heat transfer and typical convective flow speed in terms of control parameters $Ra$, $E$ and $Pr$.

\subsection{Onset regime}
\begin{figure}
    \centering
    \subfigure{
        \includegraphics[width=6.55cm,height=5.5cm]{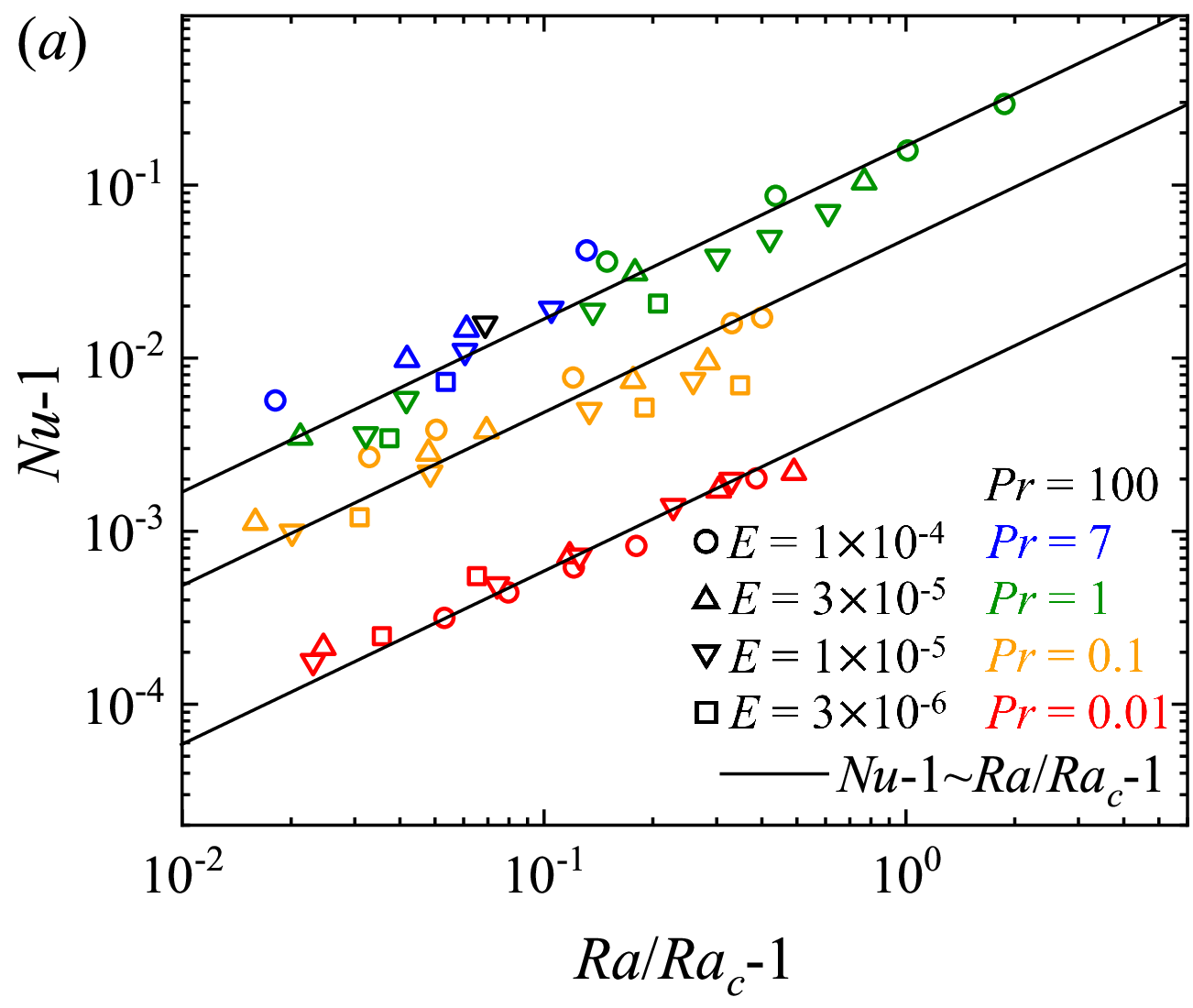}
        \label{fig:onsetnu}
    }
    \subfigure{
        \includegraphics[width=6.55cm,height=5.5cm]{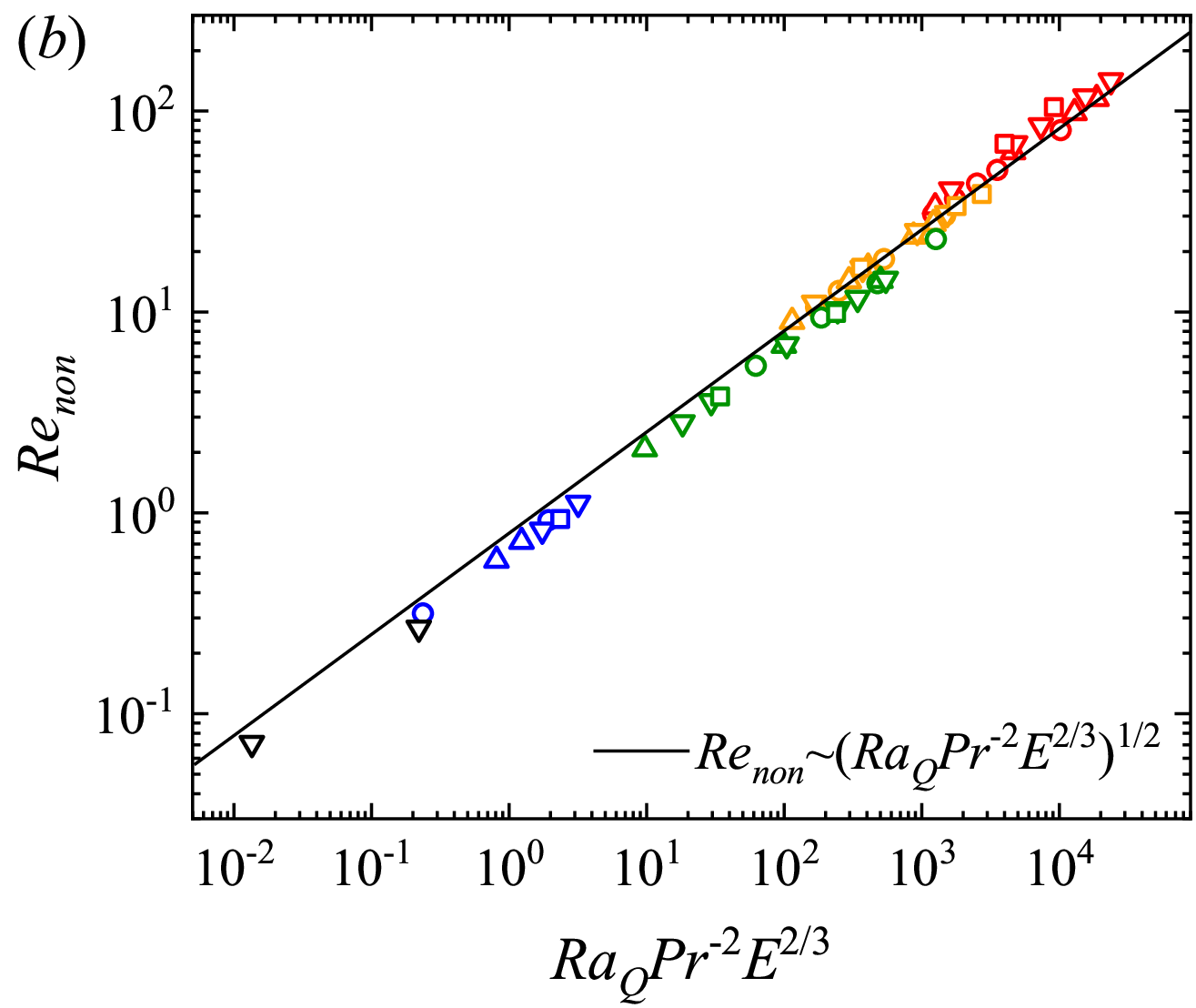}
        \label{fig:onsetre}
    }
    \caption{
($a$) Nusselt number $Nu-1$ as a function of $Ra/Ra_{c}-1$ in the onset regime. ($b$) $Re_{non}$ as a function of $Ra_{Q}\Pran^{-2}E^{2/3}$ in the onset regime. Different colors and shapes represent different values of $\Pran$ and $E$.}
\label{fig:onsetnure}
\end{figure}

We have shown the flow structures of convection onset modes are different depending on $Pr$. In this subsection, we show scaling behaviours of the heat transfer and convective flow speed near the onset, with particular attention to $Pr$-dependence. The critical Rayleigh number $Ra_{c}$ and azimuthal wavenumber $m_{c}$ of the onset mode are given in table \ref{tab:critical} for different $E$ and $\Pran$. It has been show that the convective heat transport increases linearly with $Ra/Ra_c$ near the onset \citep{busse1986convection,gillet2006quasi}  
\begin{equation}
  (Nu-1) \propto \left(\frac{Ra}{Ra_{c}}-1\right).
  \label{eq331}
\end{equation}
Figure \ref{fig:onsetnure}($a$) shows $Nu-1$ as a function of $Ra/Ra_{c}-1$ for the cases in the onset regime. We see that $Nu-1$ is linearly proportional to $Ra/Ra_c-1$ as predicted, but the prefactor of the scaling (\ref{eq331}) depends on $\Pran$, which reflects the $\Pran$-dependence of the convection onset.
For numerical models with $\Pran \ge 1$, the prefactor is similar for different $\Pran$ but weakly depends on $E$, corresponding to the viscous convection mode. At low $\Pran<1$, 
the onset modes become spiralling columnar structures that fill the whole domain outside the tangent cylinder (figure \ref{fig:3dsmallPr}($a$)), corresponding to transitional mode between viscous convection and inertial convection \citep{Zhang1992}. As the onset mode and critical $Ra_c$ depends on $\Pran$ in the transitional convection, the prefactor in the scaling (\ref{eq331}) also highly depends on $\Pran$ when $\Pran<1$. Our numerical models do not reach sufficiently low $\Pran$ to show the purely inertial convection in which the onset of convection is in the form of inertial modes \citep{zhang1994coupling}. 

For the convective flow speed $Re_{non}$ near the onset, a scaling based on the Visco-Archimedean-Coriolis (VAC) force balance has been proposed 
\citep{aubert2001systematic,king2013flow,king2013scaling}
\begin{equation}
  Re_{non}\sim Ra_{Q}^{1/2}\Pran^{-1}E^{1/3}, 
  \label{eq332}
\end{equation}
where $Ra_Q=(Nu-1)Ra$ is the flux-based Rayleigh number. 
The above VAC scaling can also be derived from the balance between the viscous dissipation rate and work done by the buoyancy force \citep{gastine2016scaling}.  
Figure \ref{fig:onsetnure}($b$) shows $Re_{non}$ versus $Ra_{Q}\Pran^{-2}E^{2/3}$ for all cases near the onset. We can see that numerical results with different $\Pran$ well fit a unified VAC scaling (\ref{eq332}). We note that the VAC scaling is given in terms of $Ra_Q$ which implicitly carries the $\Pran$-dependence of $Nu$. The VAC scaling has been confirmed by previous numerical simulations at $\Pran$ around the unity \citep{gastine2016scaling,long2020scaling}. At low $\Pran$ in our numerical model, convection near the onset presumably should be the transitional convection \citep{Zhang1992}, in which the effect of inertial force is gradually becoming apparent. However, the VAC scaling tends to be still valid for our numerical models at low $\Pran$. This means that the $\Pran$ we calculated is still not small enough to reach the inertial onset regime where the inertial force plays a dominant role.

In summary for the onset regime, both the heat transfer and typical convective flow speed can be explained by previous theoretical predictions. The prefactor of the heat transfer scaling depends on $\Pran$, which reflects the the $\Pran$-dependence of the convection onset mode.

\subsection{Multiple modes regime} \label{sec:MM}

As we increase $Ra$, the convective flow remains laminar but exhibits several modes interactions (figures \ref{fig:3dlargePr}($b$) and \ref{fig:3dsmallPr}($b$)). We refer to such flow patterns  as the multiple modes (MM) regime which corresponds to a transition from simple convection onset modes to more complex convection modes. As this regime exists only in a narrow parameter range (see figure \ref{fig:all}), we do not analyze systematic scaling behaviours but focus on the characteristics of multiple modes interaction in the MM regime. 

\begin{figure}
    \centering
    \subfigure{
        \includegraphics[width=6.55cm,height=5.5cm]{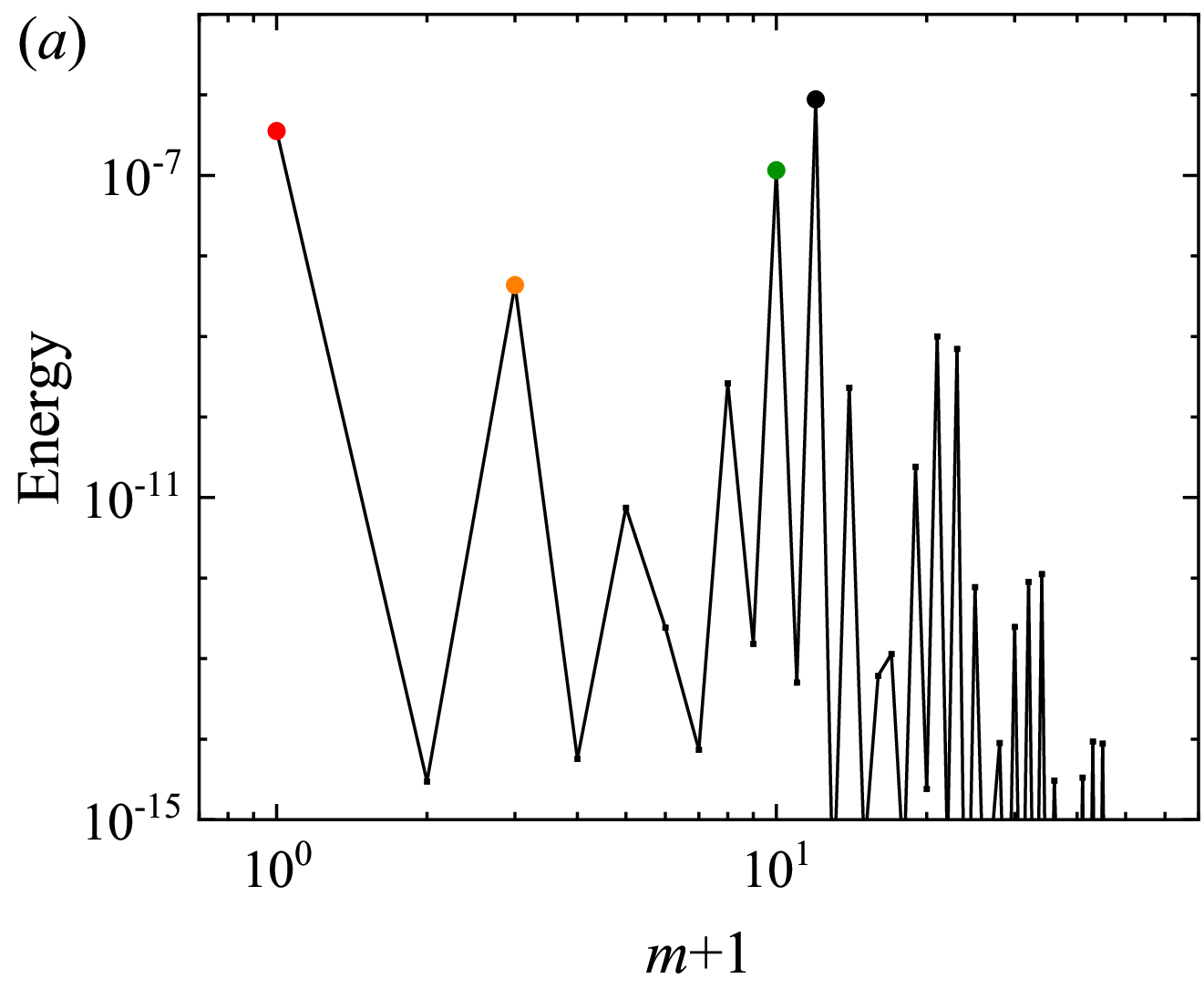}
        \label{fig:tpr01s}
    }
    \subfigure{
        \includegraphics[width=6.55cm,height=5.5cm]{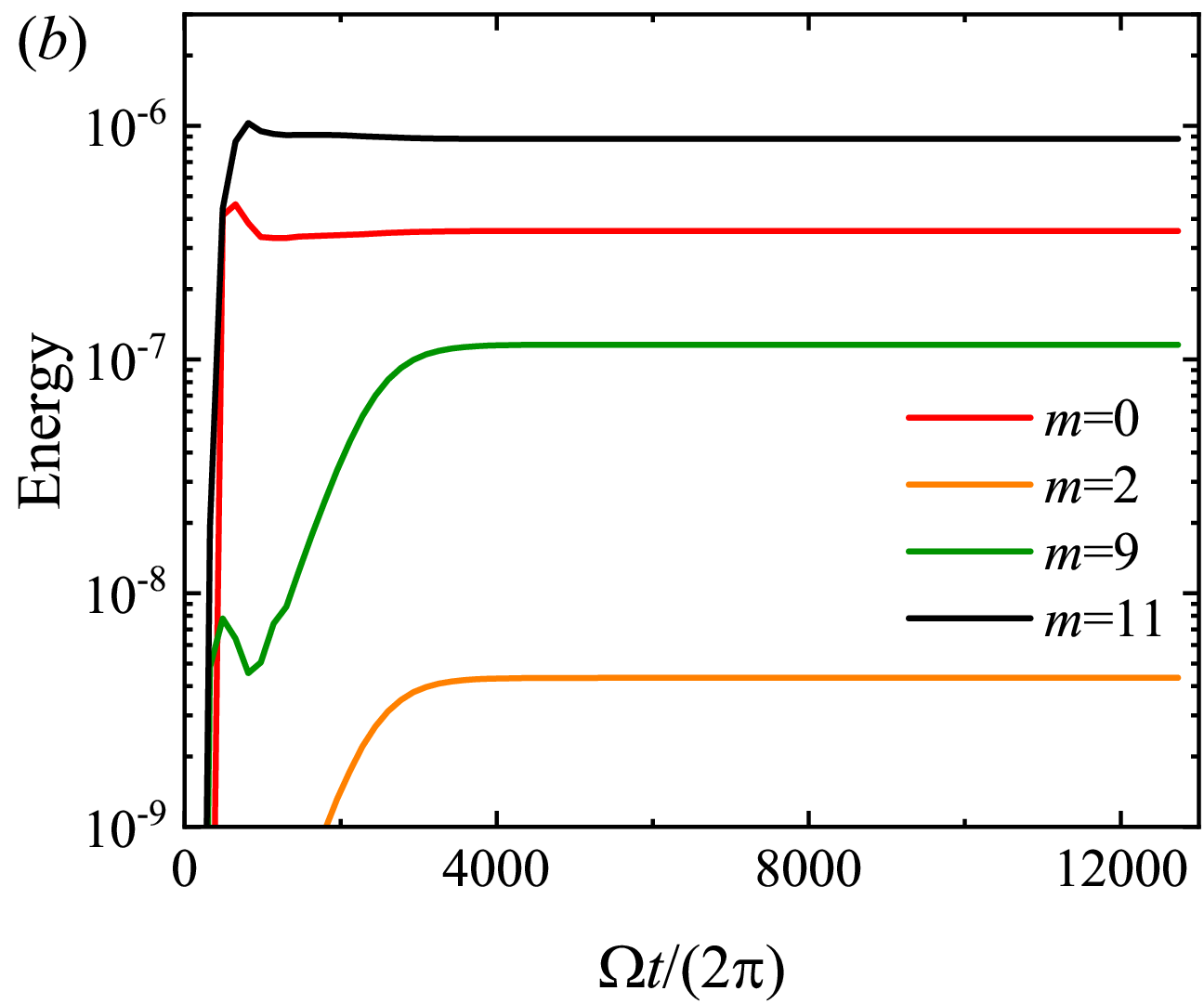}
        \label{fig:tpr01m}
    }
    \subfigure{
        \includegraphics[width=6.55cm,height=5.5cm]{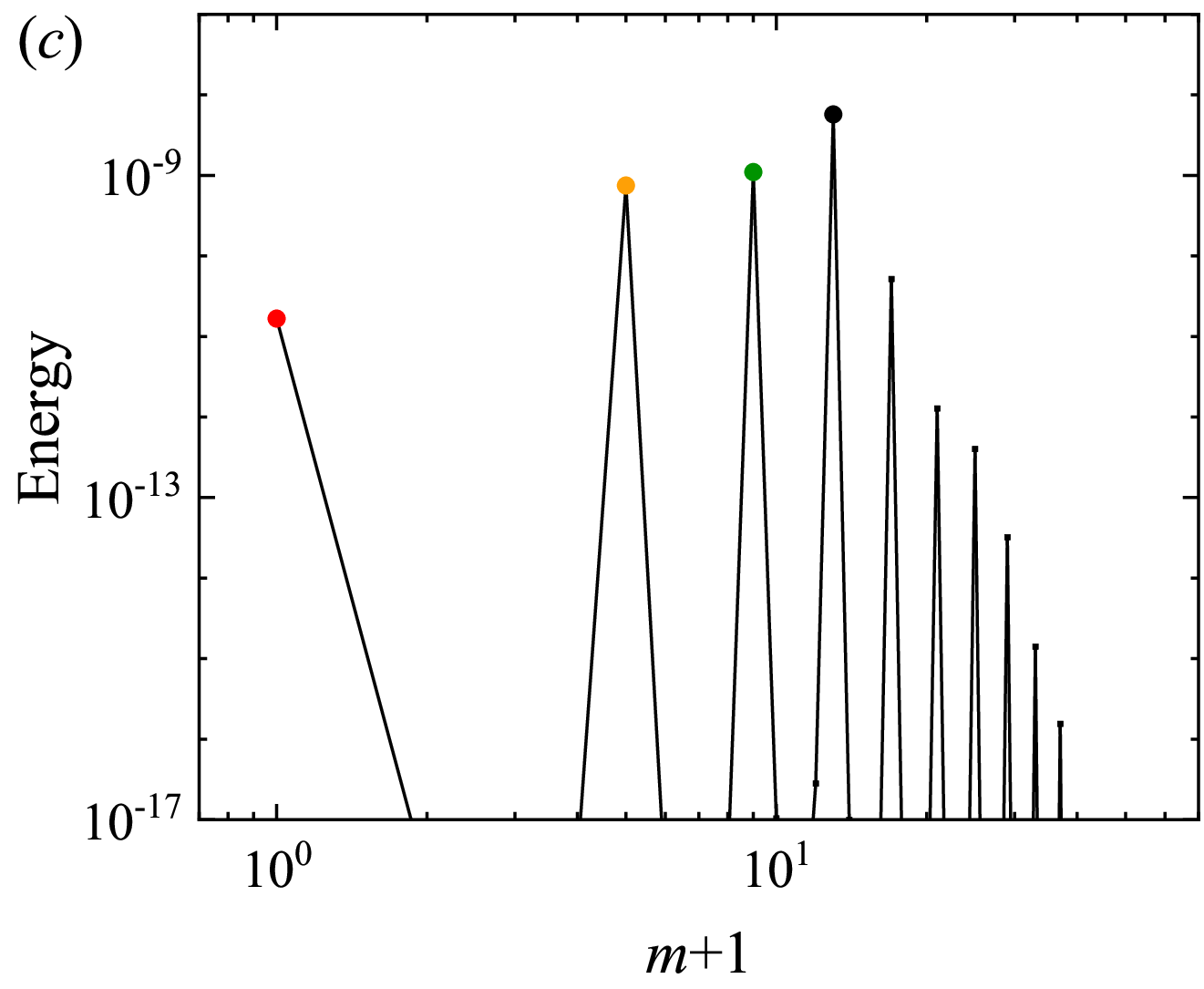}
        \label{fig:tpr7s}
    }
    \subfigure{
        \includegraphics[width=6.55cm,height=5.5cm]{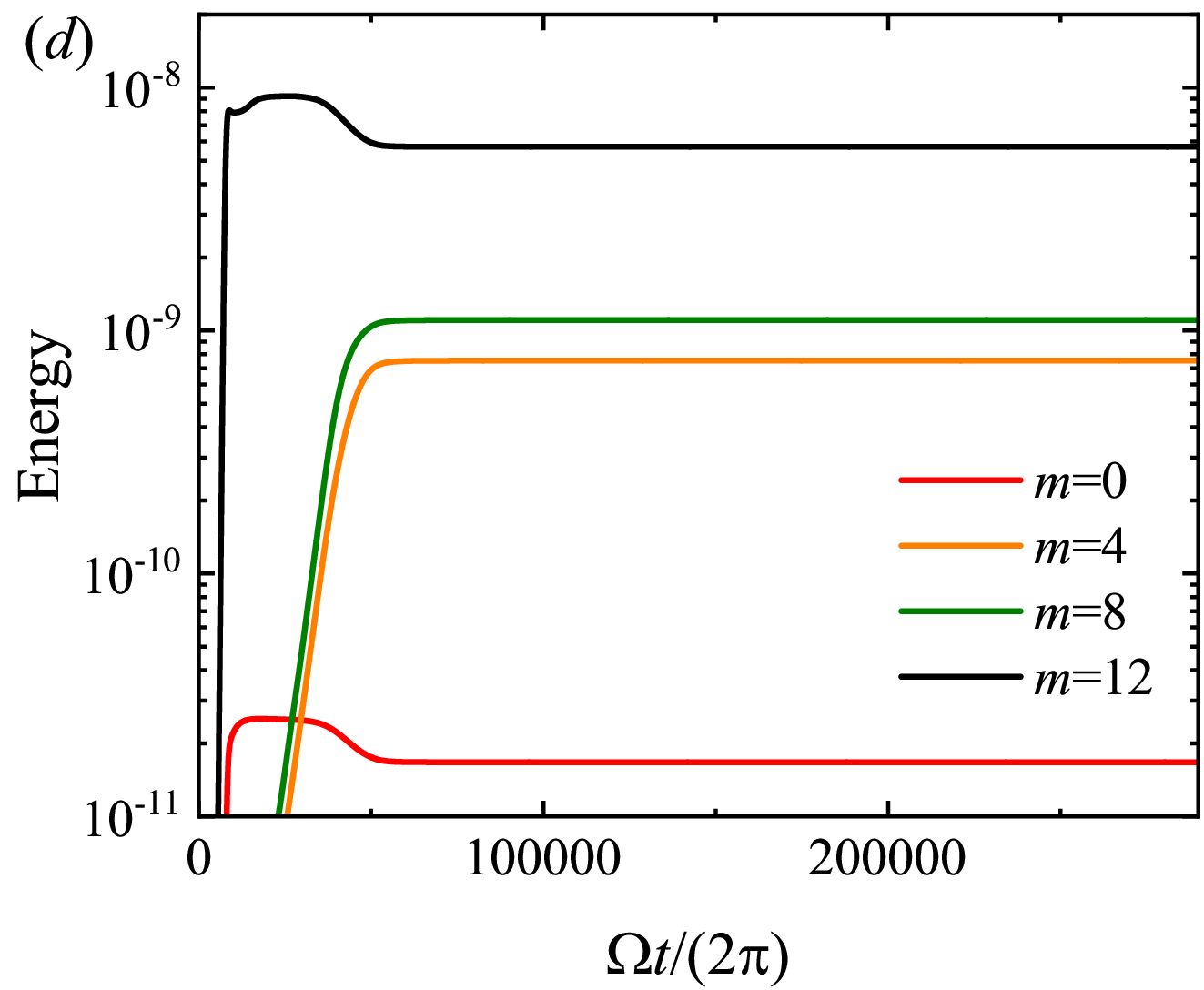}
        \label{fig:tpr7m}
    }
    \caption{Energy spectra of two cases in the MM regime with possible triadic resonances. 
    ($a$-$b$) the case with $E=1\times10^{-5}$, $\Pran=0.1$ and $Ra=5.1\times10^{6}$; ($c$-$d$) the case with $E=3\times10^{-5}$, $\Pran=7$ and $Ra=4.67\times10^{6}$.  Left panel: Time-averaged kinetic energy spectrum as a function of the azimuthal wavenumber $m$.  Right panel: Time evolution of the kinetic energy contained in different $m$ components indicated by circles in the left panel.}
    \label{fig:tpr01and07}
\end{figure}

Previous numerical simulations \citep{Horn2017,Lam2018} and laboratory experiments \citep{Aurnou2018} using liquid gallium ($\Pran \approx 0.025$) observed multiple modes interactions in rotating convection. More recently, \citet{lin2021triadic} revealed that the multiple mode interaction take place in the form of triadic resonances in spherical rotating convection at low $\Pran\le 0.01$. Triadic resonance is a generic instability mechanism in rotating fluids \citep{kerswell2002elliptical,le2015flows}, in which a primary inertial mode with azimuthal wavenumber $m_0$ and frequency $\omega_0$ can excite a pair of unstable inertial modes with wavenumbers $m_1, m_2$ and frequencies $\omega_1$, $\omega_2$ matching the resonance conditions:
\begin{equation}
  \omega_0=\omega_1 \pm \omega_2, \quad m_0=m_1 \pm m_2.
  \label{eq333}
\end{equation}

In this study we find that triadic resonances can take place at both small and moderate $\Pran$.  
Figures \ref{fig:tpr01and07} shows time-averaged energy spectra (left panel) and time evolution of the kinetic energy contained in the four dominant components (right panel)  for two cases with $\Pran=0.1$ (top panel) and $\Pran=7$ (bottom panel). For the case with $\Pran=0.1$, the four highest peaks in the $m$-spectrum correspond to $m=11, 0, 9, 2$ respectively. The $m=11$ component represents the primary convective mode, while the $m=0$ component can be attributed to the mean flow generated by the nonlinear interaction of the primary mode \citep{zhang2017theory}. As the energy in the  primary mode saturated, two other components with $m=2$ and $m=9$ that satisfy the triadic resonance condition start to exponentially grow and then saturate. This is a typical behaviour of the triadic resonance \citep{lin2021triadic} and suggests that two secondary modes are excited through the triadic resonance with the primary mode.  
The case with $\Pran=7$ in the bottom panel exhibits similar behaviours but with the primary mode of $m=12$ and two secondary modes of $m=4$ and $m=8$.
 We should note that triadic resonances at moderate $\Pran$ should be due to nonlinear interactions of three thermal Rossby waves rather than purely inertial waves at very low $\Pran$ rotating convection \citep{lin2021triadic} or in the mechanical driven rotating flows \citep{le2015flows}.

Our numerical results, in line with the early work of \cite{lin2021triadic}, suggest that the triadic resonance may provide a generic mechanism of the transition from the single onset mode to multiple modes coexisting in rotating convection with both small and moderate $\Pran$. Further increasing $Ra$  would excite more and more unstable modes and eventually lead to turbulent convection which will be discussed in the following subsection.

\subsection{Geostrophic turbulence regime} \label{subset:GT}

\begin{figure}
    \centering
    \subfigure{
        \includegraphics[width=6.55cm,height=5.5cm]{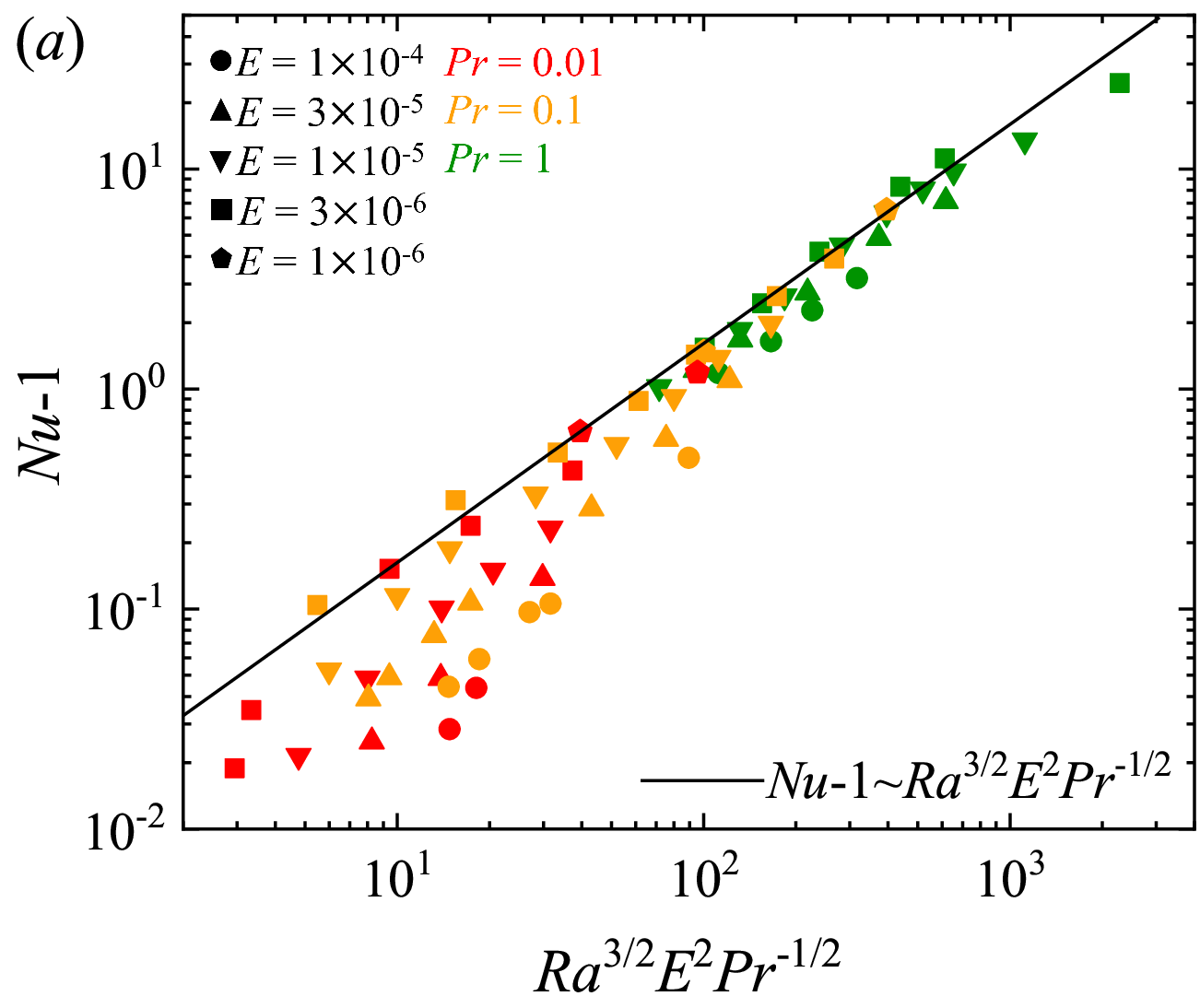}
        \label{fig:rrinu}
    }
    \subfigure{
        \includegraphics[width=6.55cm,height=5.5cm]{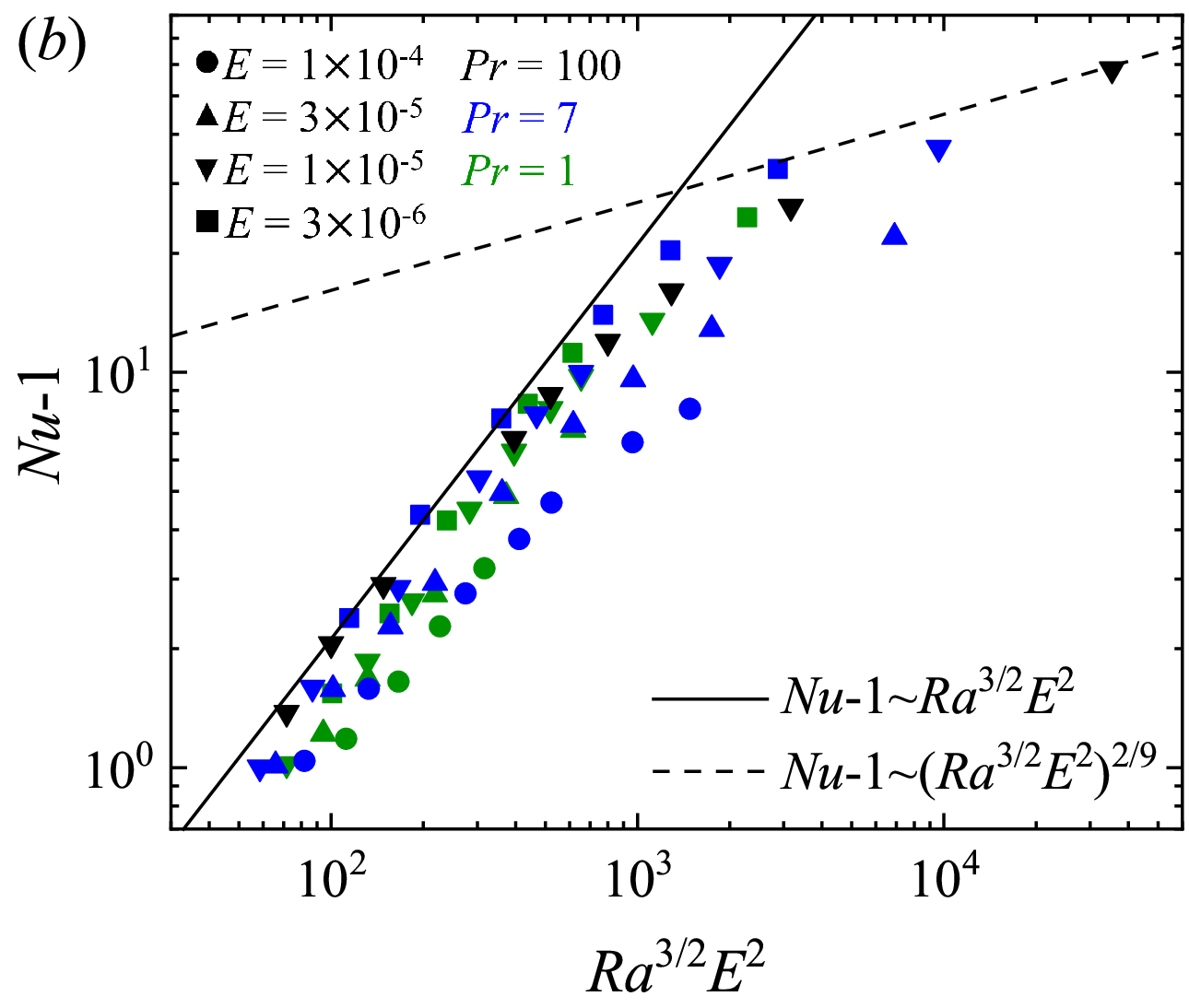}
        \label{fig:rrvnu}
    }
    \caption{Nusselt number versus control parameters in the GT regime. ($a$) $Nu-1$ as a function of $Ra^{3/2} E^2\Pran^{-1/2}$ for cases with $\Pran\le 1$. ($b$) $Nu-1$ as a function of $Ra^{3/2} E^2$ for cases with  $\Pran\ge 1$. Different shapes and colors represent different $E$ and $\Pran$ as in Figure \ref{fig:all}.}
    \label{fig:gtnu}
\end{figure}

Geostrophic turbulence is thought to be the most relevant regime to convection in rapidly rotating stars and planets. Both the flow morphology and scaling behaviours of heat transfer and momentum transport have been extensively discussed in previous studies \citep[e.g.][]{aurnou2015rotating}. In this section, we focus on the scaling behaviours of heat transfer $Nu$ and the convective flow speed $Re_{non}$ at different $\Pran$.    

In the GT regime, the well-known diffusion-free scalings have been widely discussed and compared with numerical simulations and laboratory experiments \citep{Cheng2018,hawkins2023laboratory}. The diffusion-free scalings (also called inertial scalings or CIA scalings) were derived using different approaches and arguments in the literature \citep[e.g.][]{stevenson1979turbulent,aubert2001systematic,gillet2006quasi,julien2012heat,barker2014theory,Aurnou2020}, which are  essentially based on the CIA force balance and the mixing length theory, leading to scalings independent of the fluid viscosity and thermal diffusivity. These scalings predict that the efficiency of convective heat transfer follows 
\begin{equation}
  (Nu-1)\sim Ra^{3/2}E^{2}\Pran^{-1/2},
  \label{eq334}
\end{equation}
and the convective flow speeds follows
\begin{equation}
  Re_{non}\sim (Ra_{Q}\Pran^{-2}E^{1/2})^{2/5},
  \label{eq336}
\end{equation}
where $Ra_Q=(Nu-1)Ra$ is the flux-based Rayleigh number. Previous studies found that the scalings (\ref{eq334}-\ref{eq336}) can fit numerical and experimental data in certain parameter regime, but most of numerical simulations set $\Pran$ around the unity and laboratory experiments usually use water ($\Pran\approx 7$) as working fluids (but see other rotating convection experiments with different $\Pran$ \citep[e.g.][]{Aurnou2018,Abbate2023}). Systematic test of the diffusion-free scalings over a wide range of $\Pran$ seems to be still lacking. Here we compare the diffusion-free scalings with our numerical models over a wide range of $\Pran$ ($10^{-2}\le \Pran \le 10^2$).

Figure \ref{fig:gtnu} shows $Nu-1$ as a function of the control parameters $Ra$, $E$ and $\Pran$ in the GT regime. As we already noticed from figure \ref{fig:all} that a single scaling is not able to reconcile numerical results with different $\Pran$ because $Nu-1$ highly depends on $\Pran$ when $\Pran \le 1$ and becomes nearly independent of $\Pran$ when $\Pran \ge 1$. Therefore, we plot $Nu-1$ separately for cases with $\Pran \le 1$  and $\Pran \ge 1$ in figure \ref{fig:gtnu}. We can see from figure \ref{fig:gtnu}($a$) that the diffusion-free scaling (\ref{eq334}) matches numerical data reasonably well with $\Pran \le 1$ and $(Nu-1)\ge 1$, except some data points in the left-bottom which correspond to cases with $(Nu-1)<1$. In these cases, the convective flow becomes chaotic but $Nu$ remains small because of efficient thermal conduction at small $\Pran$. The diffusion-free scaling is expected to be valid in the asymptotic regime of $Nu\gg1$. However, it is a huge numerical challenge to achieve large $Nu$ but remain in GT regime with small $\Pran$, as this would require reducing $E$ and increasing $Ra$ meanwhile. So we find that the diffusion-free scaling (\ref{eq334})  for the convective heat transfer tends to be valid when $\Pran \le 1$ and $(Nu-1)\ge 1$ based on our numerically accessible models. 

\begin{figure}
    \centering
    \subfigure{
        \includegraphics[width=6.55cm,height=5.5cm]{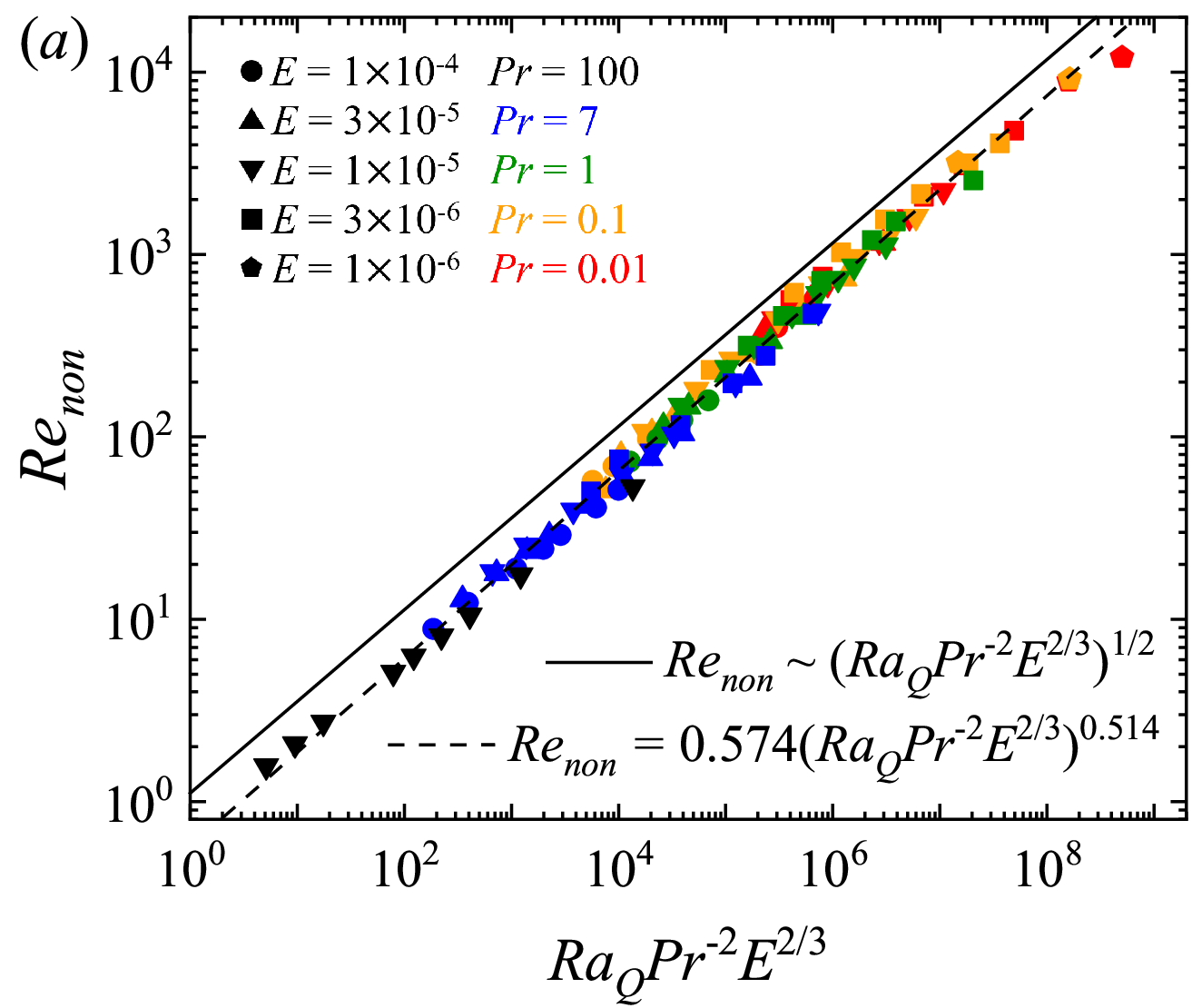}
        \label{fig:vac}
    }
    \subfigure{
        \includegraphics[width=6.55cm,height=5.5cm]{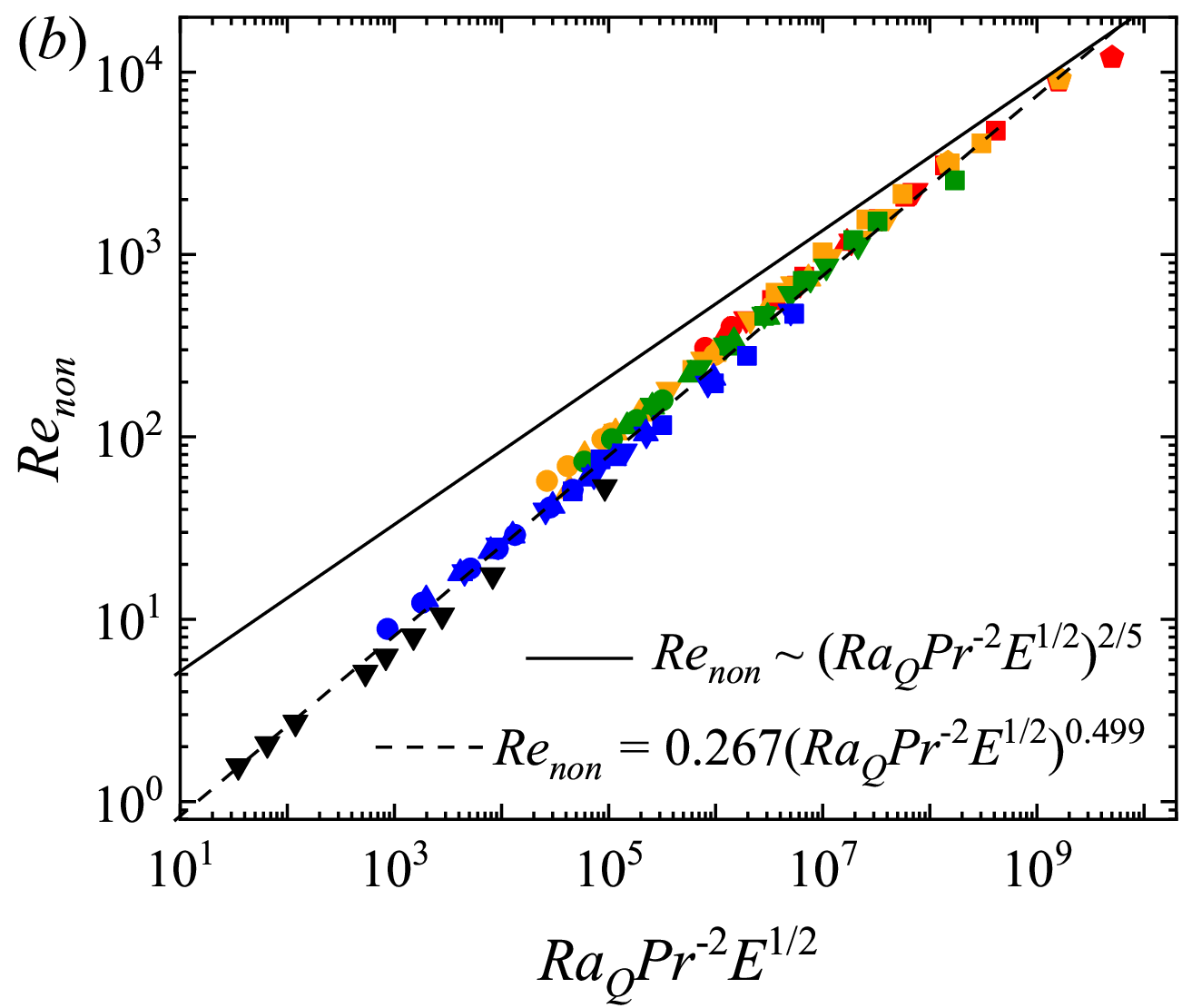}
        \label{fig:cia}
    }
  \caption{Non-zonal Reynolds number $Re_{non}$ versus control parameters in the GT regime. ($a$) $Re_{non}$ as a function of $Ra_{Q}\Pran^{-2}E^{2/3}$. The solid black line corresponds to the VAC scaling (\ref{eq332}). ($b$) $Re_{non}$ as a function of $Ra_{Q}\Pran^{-2}E^{1/2}$. The solid black line corresponds to the CIA scaling (\ref{eq336}). The dashed lines corresponds to the least-square fit to the data.}
\label{fig:rrre}
\end{figure}

As $Nu$ becomes nearly independent of $\Pran$ when  $\Pran\ge 1$ (in figure \ref{fig:all}), figure \ref{fig:gtnu}($b$) plots $Nu-1$ as a function of $Ra^{3/2}E^2$ that without $\Pran$-dependence for cases with $\Pran\ge 1$ in the GT regime. Here we basically drop the $\Pran$-dependency and keep the same power law for $Ra$ and $E$ as in the diffusion-free scaling (\ref{eq334}) for reference. We can see that $Nu-1$ does not show obvious dependence on $\Pran$ for fixed $Ra$ and $E$. The almost independence of $Nu$ on $\Pran$ at moderate and large $\Pran$ is in line with previous studies on non-rotating convection \citep{verzicco1999prandtl,Li2021}. 
The scattering of data points from the solid line in figure \ref{fig:gtnu}($b$)  is mainly due to different Ekman numbers (different shapes in the plot), suggesting that the power law $E^2$ is also not suitable for the $E$-dependence. For fixed $E$ and $\Pran$, we see that $Nu-1$ is proportional to $Ra^{3/2}$ mostly but tends to deviate from the power law $Ra^{3/2}$ and approach to the non-rotating power law $Ra^{1/3}$ at large $Ra$, although the convection remains rotationally dominated based on the criterion of $Ro_{\ell} <0.1$. We will discuss the transition from rotating turbulent convection to non-rotating convection in section \ref{subsec:WR}. In short, figure \ref{fig:gtnu}($b$) shows that the widely used diffusion-free scaling (\ref{eq334}) for the convective heat transfer does not fit our numerical models with $\Pran \ge 1 $ in the GT regime, which is also shown by recent laboratory experiments with moderate to high $\Pran$ \citep{Abbate2023}.

We now turn to examine the scaling behaviour of the convective flow speed measured by the non-zonal Reynolds number $Re_{non}$. Figure \ref{fig:rrre}($a$) and \ref{fig:rrre}($b$) shows comparisons of $Re_{non}$ from numerical models in the GT regime with the VAC scaling and the CIA scaling respectively. By comparing the results of least-squares fitting, we found that our numerical results more closely collapse on the VAC scaling. However, we note that the flow speed tends to approach the CIA scaling at low $\Pran$ and large $Re_{non}$ ($\gtrsim 10^3$). This implies that the viscosity plays a non-negligible role in most of our numerical models. It is interesting that convective velocities roughly follow a unified scaling, despite the very different scaling behaviours for the heat transfer.    
However, we should mention that both the VAC and CIA scalings are given in terms of the flux-based Rayleigh number $Ra_Q=(Nu-1)Ra$, which already takes into account different scaling behaviours of $Nu-1$. The scaling behaviour of the convective velocities in our numerical simulations is also in agreement with recent RRBC laboratory experiments at different $\Pran$ \citep{Abbate2023}. 

In summary for the GT regime, both heat transfer and convective velocities asymptotically approach the diffusion-free scalings at low $\Pran\le 1$, suggesting that both the global heat transfer and convective flows are controlled by inviscid dynamics in the bulk. At large $\Pran>1$, the efficiency of heat transfer becomes nearly independent of $\Pran$ and approaches the non-rotating scaling at large $Ra$, while the convective velocities closely follow the scaling based on the VAC force balance. The scaling behaviours at large $\Pran$, in line with experimental results of \cite{Abbate2023}, suggest that the heat transfer is controlled by the boundary layers, whereas the typical flow speeds are controlled by the interior force balance in currently accessible numerical models.     

\subsection{Weakly rotating regime} \label{subsec:WR}

\begin{figure}
    \centering
    \subfigure{
        \includegraphics[width=6.55cm,height=5.5cm]{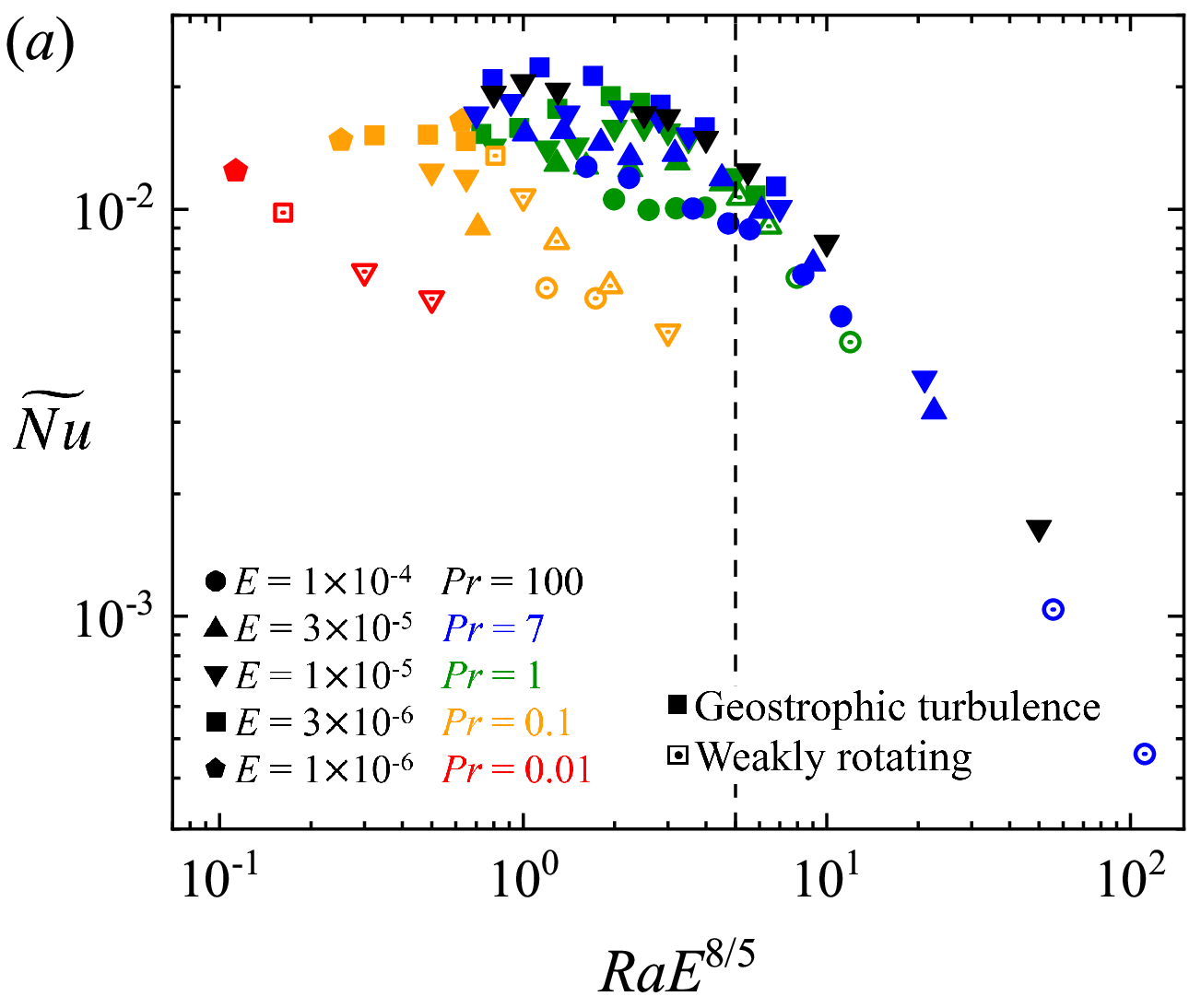}
    }
    \subfigure{
        \includegraphics[width=6.55cm,height=5.5cm]{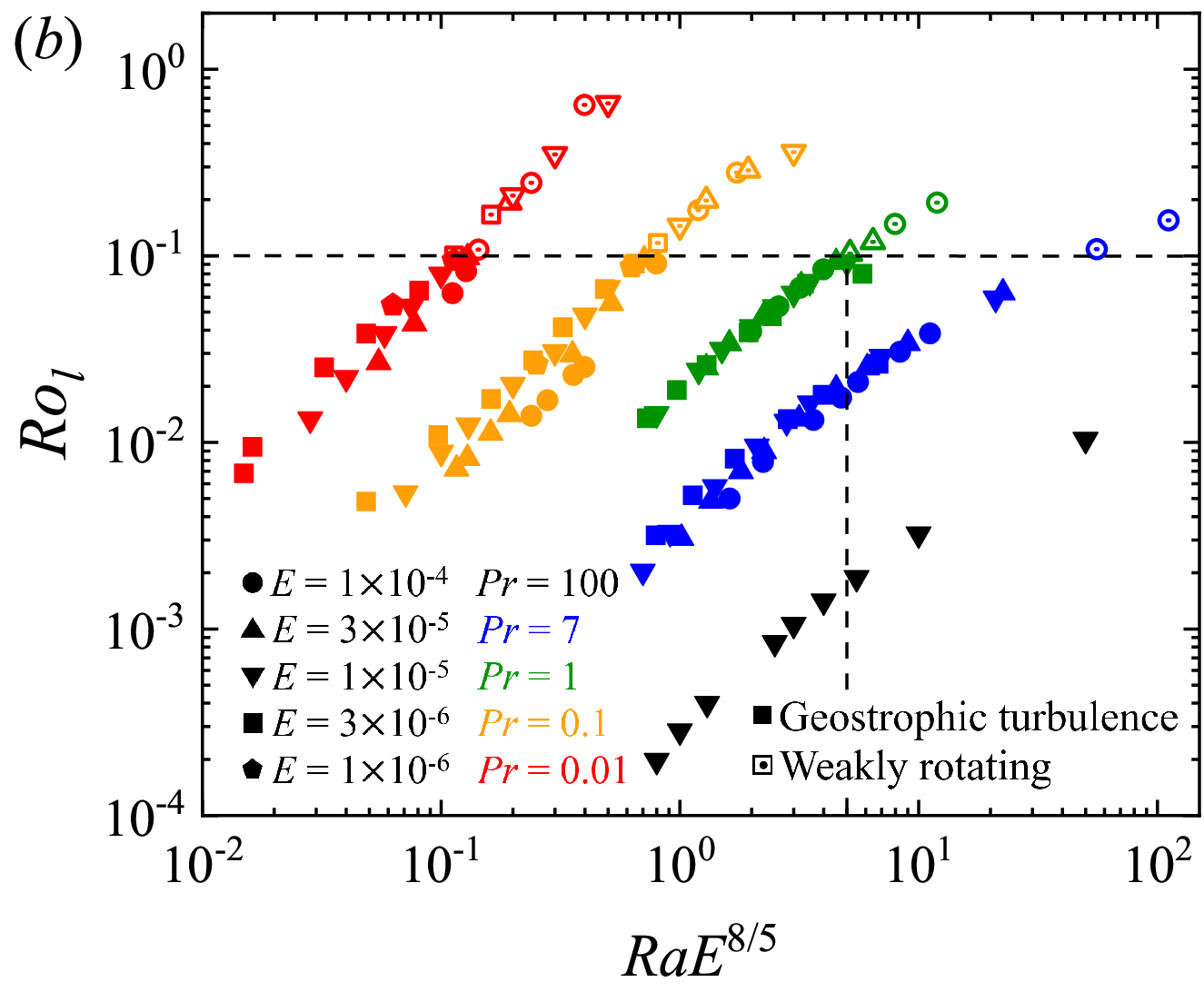}
    }
    \subfigure{
        \includegraphics[width=6.55cm,height=5.5cm]{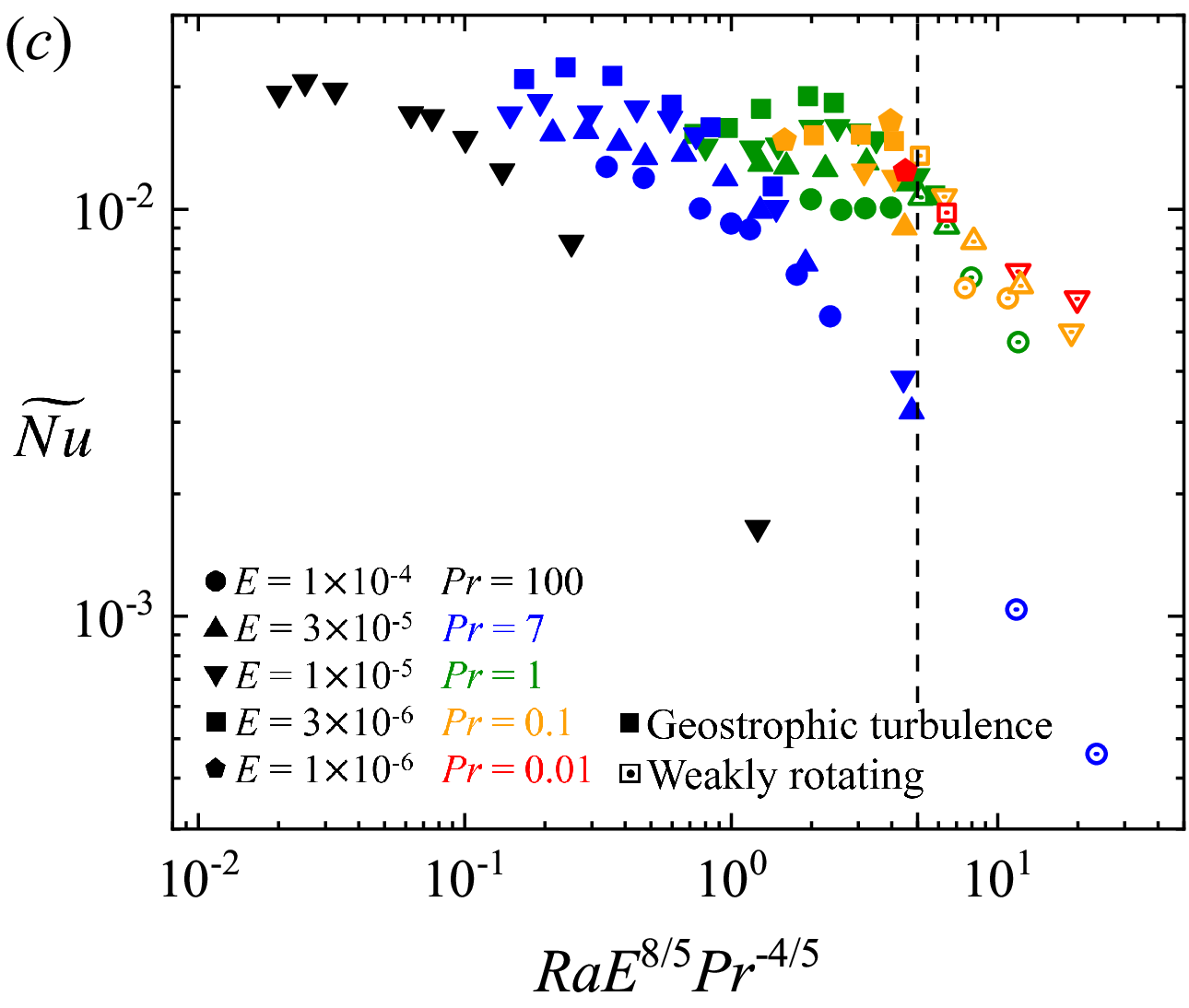}
    }
    \subfigure{
        \includegraphics[width=6.55cm,height=5.5cm]{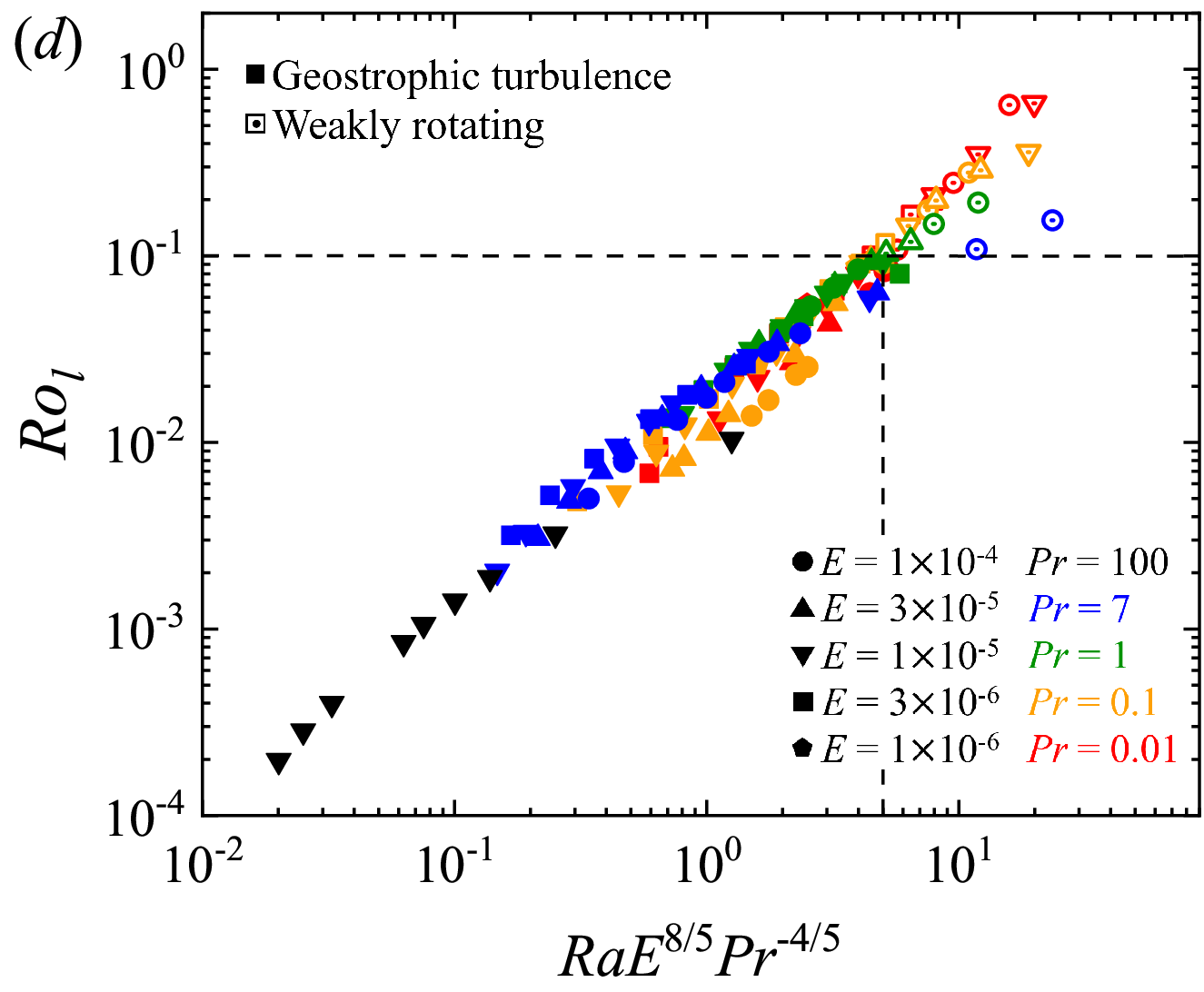}
    }
    \caption{Rescaled Nusselt number $\widetilde{Nu}$ and local Rossby number $Ro_\ell$ in the GT and WR regimes. ($a$) $\widetilde{Nu}$ and ($b$) $Ro_\ell$ as a function of $RaE^{8/5}$; ($c$)  $\widetilde{Nu}$ and ($d$)  $Ro_\ell$ as a function of $RaE^{8/5}Pr^{-4/5}$. Data points with $Nu<2$ are excluded in ($a$) and ($c$).  
    Symbols have the same definitions as those in figure \ref{fig:all}.}
    \label{fig:wr}
\end{figure}
Further increasing $Ra$, the convective flows become less geostrophic due to the weakening rotational constraint (figures \ref{fig:3dlargePr}($d$) and \ref{fig:3dsmallPr}($d$)). At sufficiently large $Ra$, strong buoyancy force would break rotational constraints and lead to non-rotating convection. However, the transition from rotating to non-rotating convection does not mean an abrupt regime change. It is not straightforward to quantitatively define when the convection is rotationally dominated.  
The regime changes from rotating to  non-rotating convection have been extensively discussed and several transition criteria were proposed and tested \citep[e.g.][]{King2009,julien2012heat,gastine2016scaling,long2020scaling}, yet no consensus has been reached.
As introduced in section  \ref{subsec:OV}, we use the local Rossby number $Ro_{\ell} \le 0.1$ as a tentative criterion for the rotation-dominated convection and thus set cases with $Ro_{\ell} > 0.1$ as the weakly rotating regime. In this section we examine changes of the Nusselt number and local Rossby number as increasing $Ra$ from GT to WR regimes. As we shall show the transitional behaviours are rather complicated and it is impractical to define a single transition criteria that can reconcile both the heat transfer and flow morphology with different $\Pran$.

In order to monitor the behaviour changes of the heat transfer, we define a rescaled Nusselt number $\widetilde{Nu}$:
\begin{equation}
    \widetilde{Nu} =\left\{ 
    \begin{aligned}
        \frac{Nu-1}{Ra^{3/2}E^{2}} & \quad \mathrm{when}\, \Pran \ge 1, \\
        &\\
        \frac{Nu-1}{Ra^{3/2}E^{2}\Pran^{-1/2}} & \quad \mathrm{when}\, \Pran \le 1,
    \end{aligned}
     \right. 
\end{equation}
which takes into account different scalings of $Nu-1$ at different $\Pran$ in the GT regime as shown in figure \ref{fig:gtnu}. The rescaled $\widetilde{Nu}$ should be flat  in the rotation-dominated convection and would start to drop as increasing $Ra$ to the weakly rotating regime. The turning point is usually seen as a sign for the transition from rotation to non-rotating convection \citep{long2020scaling}.

Figure \ref{fig:wr}($a-b$) show the rescaled Nusselt number $\widetilde{Nu}$ and the local Rossby number $Ro_\ell$ as a function of $RaE^{8/5}$, which was proposed to be a key control parameter for the transition \citep{julien2012heat}.  We can see that the heat transfer behaviours indeed change at around $RaE^{8/5}\approx5$ for data points with $\Pran\ge1$, but this criteria does not reconcile data points with low $\Pran$. It is interesting that $RaE^{8/5}\approx5$  also corresponds to $Ro_{\ell}\approx0.1$ at $\Pran=1$. However, the $Ro_{\ell}$ at high $\Pran$ (blue and black symbols in figure \ref{fig:wr}($b$)) is still much smaller than 0.1 even when $RaE^{8/5}>5$. This means that the convective flow remains quite geostrophic in the bulk but the heat transfer already approaches the non-rotating scaling at high $\Pran$. Therefore, the transitional criterion based on the heat transfer and based on the flow morphology would be quite different. This again points to the scenario that the heat transfer is controlled by the boundary layers but the convective flows are controlled by the force balance in the bulk.   

Figure \ref{fig:wr}($b$) clearly shows that the $Ro_{\ell}$ depends on $\Pran$. By an empirical fitting of data, we find that the $\Pran$-dependence can be approximated by a power exponent of around 4/5. Therefore, we define a control parameter 
\begin{equation}
    Ra_G=RaE^{8/5}\Pran^{-4/5},
\end{equation}
and plot $Nu$ and $Ro_{\ell}$ as a function of $Ra_G$ in figure \ref{fig:wr}($c-d$). 
\cite{gastine2016scaling} derived a similar control parameter of $RaE^{8/5}\Pran^{-3/5}$ based on the convective Rossby number $Ro_c$ in the boundary layers. Here we find that $Ra_G=RaE^{8/5}\Pran^{-4/5}$ can better fit our numerical data of the $Ro_{\ell}$. Figure \ref{fig:wr}($d$) shows that $Ra_G\approx5$ corresponds to $Ro_{\ell}=0.1$ for all different $\Pran$. We also seen from figure \ref{fig:wr}($c$) that $Ra_G\approx5$ corresponds to the turning point of the rescaled Nusselt number $\widetilde{Nu}$ for data points with $\Pran\le 1$. This implies that the transition criteria can be unified for the heat transfer and flow morphology at low $\Pran$.   
The control parameter $Ra_G$ is not suitable to describe the transition of the heat transfer for model with $\Pran>1$ as expected. 

Combining the scaling behaviours shown in section \ref{subset:GT} and the transitional behaviours in this section, we suggest that the heat transfer is controlled by the boundary layers at large $\Pran$ and by the interior dynamics at small $\Pran$. We should mention that $\Pran=1$ corresponds a quite special parameter regime because data points can fit both $\Pran$ dependent and  $\Pran$ independent scalings. The convective flow velocities and flow morphology are mainly controlled by the force balance in the interior. If we characterise the transition from GT to WR based on the $Ro_{\ell}$ in the bulk, we find $Ra_G=RaE^{8/5}\Pran^{-4/5}$ provides a unified control parameter for the transition with all different $\Pran$ based on our numerical data.

\subsection{Mean zonal flows}
\begin{figure}
    \centering
    \subfigure{
        \includegraphics[width=6.55cm,height=5.5cm]{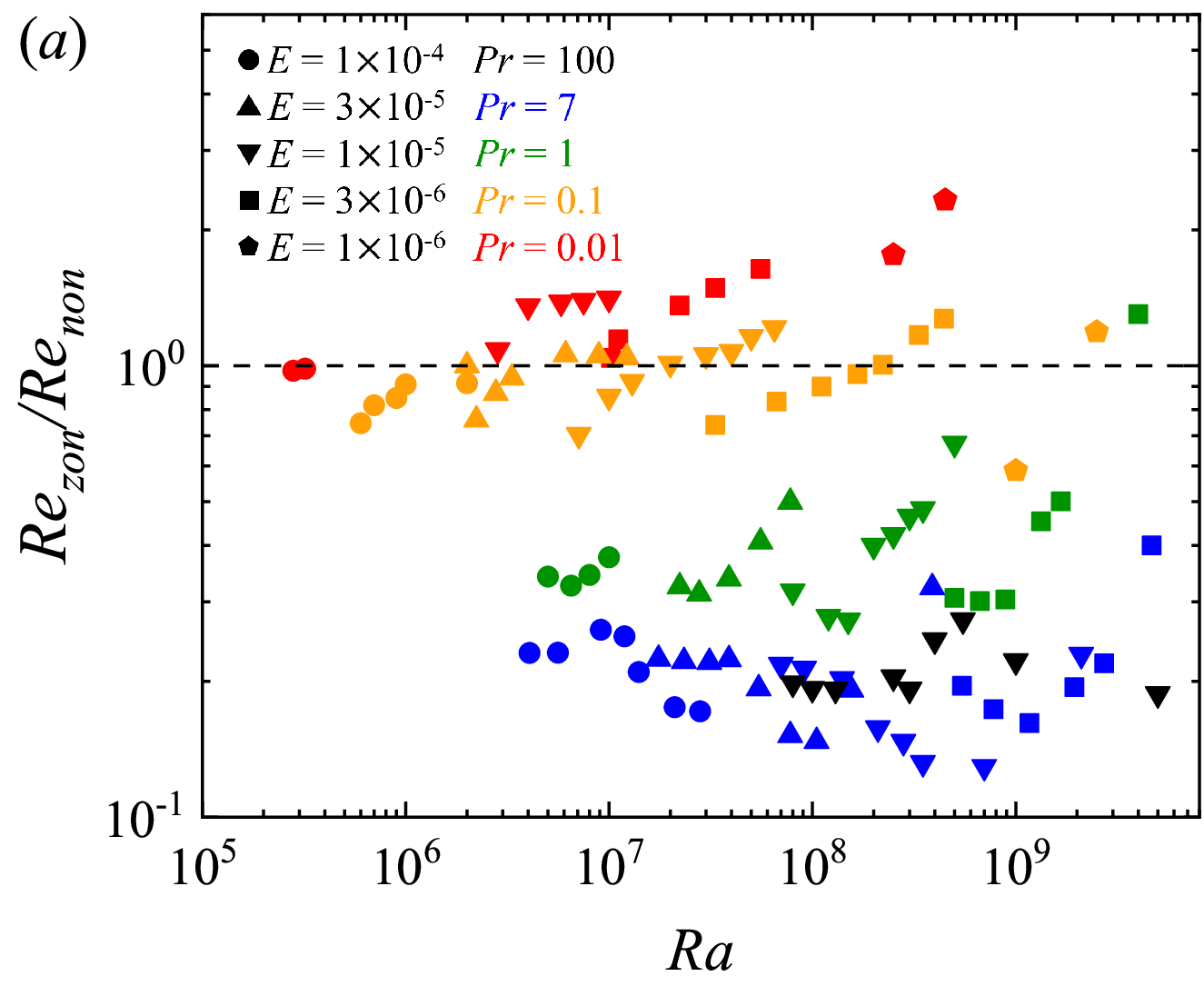}
    }
    \subfigure{
        \includegraphics[width=6.55cm,height=5.5cm]{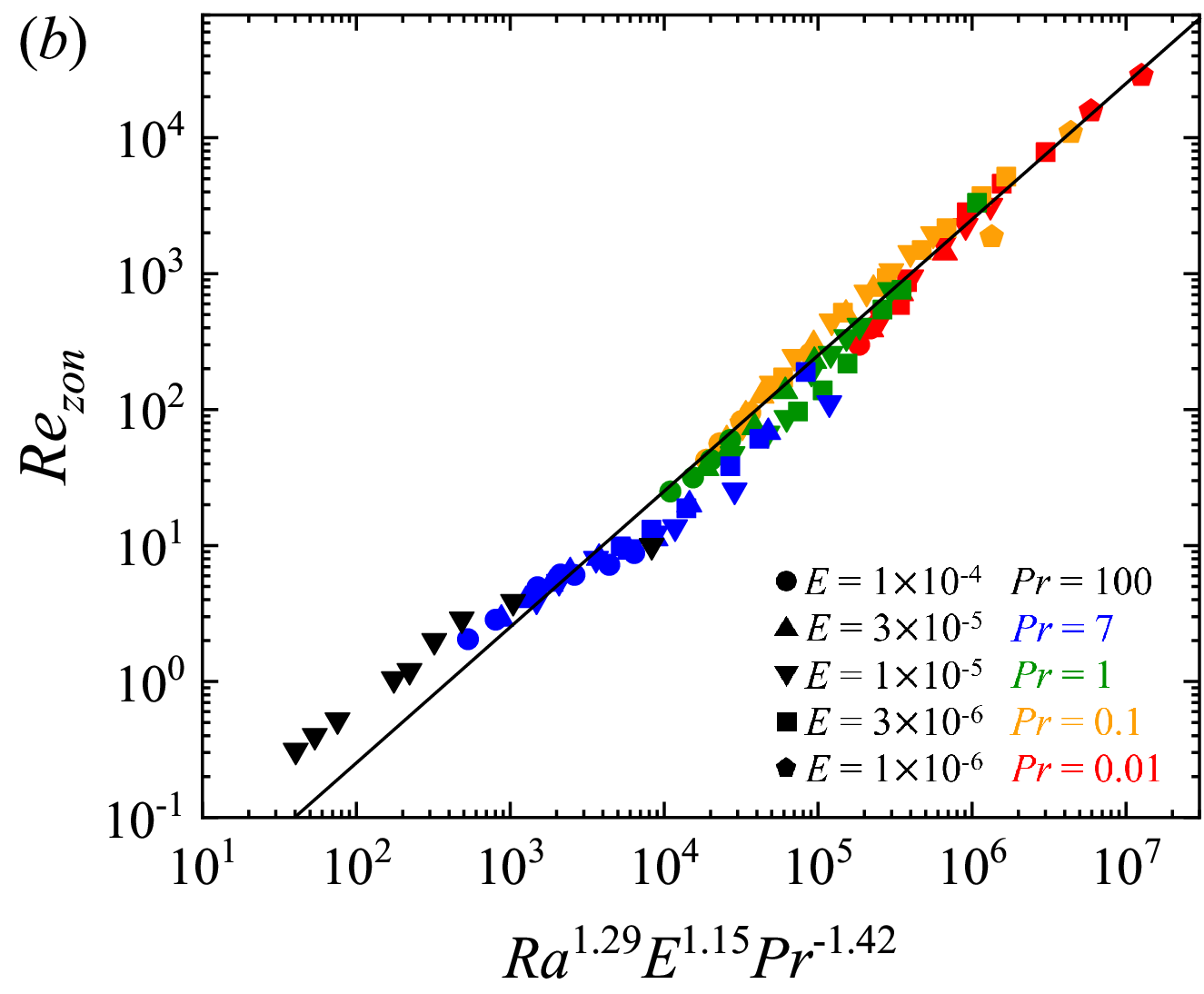}
    }
    \caption{($a$) Reynolds number ratio $Re_{zon}/Re_{non}$ as a function of $Ra$ in the geostrophic turbulence regime. ($b$) $Re_{zon}$ as a function of $Ra^{1.29}E^{1.15}\Pran^{-1.42}$ in the geostrophic turbulence regime. Symbols have the same definitions as those in figure \ref{fig:all}}.
    \label{fig:zonal}
\end{figure}

The above analysis has focused on the scaling behavior of non-zonal flows, which are related to the heat transfer. Mean zonal flows are spontaneously generated in rotating convection systems due to the nonlinear effects \citep[e.g.][]{christensen2002zonal,miyagoshi2010zonal}. In this section, we show the magnitude of zonal flows and scaling behaviours with different $\Pran$ in the GT regime where the zonal flow may be significant. Previous studies have shown that zonal flows are more readily developed at low Prandtl numbers $\Pran<1$ \citep{Zhang1992,aubert2001systematic}. The zonal flow amplitude is measured by the zonal Reynolds number $Re_{zon}$ as defined in  section \ref{sec:diagnostics}.
Figure \ref{fig:zonal}($a$) shows the ratio of $Re_{zon}/Re_{non}$ as a function of $Ra$ with different $E$ and $\Pran$ in the GT regime. 
We can see that the ratio is around or larger than unity for most of cases at $\Pran<1$, suggesting strong zonal flows are generated at low Prandtl numbers. In contrast, most of cases show the ratio $Re_{zon}/Re_{non} <1$ when $\Pran >1$. These results are in line with with previous numerical and experimental studies on zonal flows driven by rotating convection \citep{aubert2001systematic,christensen2002zonal}. It is also found that as $Ra$ increases, $Re_{zon}/Re_{non}$ in both $\Pran=1$ and 7 cases shows a significant increase, especially at relatively low Ekman number $E=3 \times 10^{-6}$. This demonstrates that at sufficiently small $E$ and large $Ra$, the zonal flows can become dominant over the non-zonal flows even at high $\Pran$. However, within the parameter interval we calculated, dominant zonal flows are more likely to be generated at low $\Pran$.

It is of great interest to see if there is any systematic dependence of the zonal flows on the control parameters. As there is no existing theoretical prediction on scaling of zonal flows in rotating turbulent convection, we made a least-square fitting of $Re_{zon}$ in the form of power law in control parameters from numerical data with $Re_{zon} \ge 100$. As shown in figure \ref{fig:zonal}($b$), the power law scaling $Re_{zon} \sim Ra^{1.29}E^{1.1}Pr^{-1.42}$ (solid black line) can well describe the data. It is found that the zonal flow strength is stronger for $\Pran \le 1$, and for $\Pran \ge 1$ the $Re_{zon}$ hardly exceeds 100 in our simulation interval.  The scaling of $\Pran^{-1.42}$ demonstrates that the strong zonal flow is more likely to be generated at $\Pran \le 1$ with fixed $Ra$ and $E$. For $\Pran \ge 1$, the strong zonal flow can also be generated by increasing $Ra$ with fixed $E$.

\section{Conclusions and outlooks}\label{sec:Conclusions}

In this study, we have constructed more than 200 numerical models of rotating convection in a spherical shell over a wide range of $\Pran$ from $10^{-2}$ to $10^{2}$, which provide a valuable dataset to investigate the scaling behaviours of rotating convection. Our numerical models are separated into four different flow regimes, namely near the onset, multiple modes interactions, geostrophic turbulence and weakly rotating regimes. We investigated the scaling behaviours  of the heat transfer and convective flow velocities in different flow regimes, with particular attention to the dependence on the $\Pran$.

Near the onset regime, the convective heat transfer is proportional to the supercriticality as predicted but the prefactor depends on the $\Pran$ due to different onset modes. The convective velocities can be well described by the VAC scaling. Multiple modes interactions can be seen as a transitional regime from laminar to turbulent convection. In this regime, we show possible evidences of triadic resonances at both low and high $\Pran$, suggesting that the triadic resonance is a generic mechanism for the transition to turbulence in rotating convection.

In the geostrophic turbulence regime, we find that a single scaling for cannot reconcile the heat transfer behaviours of numerical models with different $\Pran$. The heat transfer at low $\Pran$ tends to approach the diffusion-free scaling whereas the Nusselt numbers at high $\Pran$ become nearly independent of $\Pran$. However, the convective velocities at different $\Pran$ roughly follow a unified scaling that is in the VAC force balances, though the scaling tends to approach the CIA force balance at low $Pr$ and large $Re_{non}$.  For the mean zonal flow, we  obtain the power law scaling $Re_{zon} \sim Ra^{1.29}E^{1.1}Pr^{-1.42}$ by fitting numerical data in the GT regime.

We also find that the transition behaviours from geostrophic turbulence to weakly rotating regimes are different depending on $\Pran$ as noted in previous experimental studies \citep{King2013}. At high $\Pran$, the heat transfer already approaches the non-rotating scaling while the convective flows remain rotation-dominated based on the local Rossby number in the bulk. At low $\Pran$, transitions of the heat transfer and flow morphology take place simultaneously, both of which can be roughly determined by the control parameter $Ra_G=RaE^{8/5}\Pran^{-4/5}$ according to the empirical fitting of numerical data. In fact, we show that $Ra_G\approx 5$ provides a unified transition criterion for all different $\Pran$ if we characterise the transition merely based on the local Rossby number in the bulk. 

Both scaling behaviours and transition behaviours suggest that the heat transfer is controlled by the boundary layers whereas the convective flows are controlled by the force balance in the bulk at high $\Pran$ \citep{Abbate2023,hawkins2023laboratory}. Both numerical and experimental results show the convective velocities are close to the VAC scaling, suggesting that the viscosity still plays a non-negligible role in currently accessible numerical simulations and laboratory experiments. At low $\Pran$, it is very difficult to achieve large $Nu$ while maintain geostrophic turbulence to fully test the diffusion-free scalings. In fact, simulations at both low and high $\Pran$ pose a huge numerical challenge because of large contrasts of diffusivities. Nevertheless, our numerical models at low $\Pran$ show the trend to approach the diffusion-free scalings for both the heat transfer and convective velocities, though it require further confirmation in more extreme parameter regime. 

Finally, it is of great interests to investigate the effect of $\Pran$ on rotating convection in the presence of magnetic fields for planetary core dynamics. Meanwhile, spherical shell rotating convection is latitudinally dependent, as reported by \cite{wang2021diffusion,gastine2023latitudinal}. In this study, we considered only the global averaged scalings which represents the overall dynamics in the system. It would be interesting to  to explore the scaling of different latitudes under different $\Pran$ to complement our study.



\backsection[Acknowledgements]{}
Numerical simulations made use of the open-source code XSHELLS, which is freely available at \url{https://nschaeff.bitbucket.io/xshells}.
We thank Nathanaël Schaeffer for his assistance in setting up and running the numerical code. The critical Rayleight numbers and wavenumbers for the onset of convection were calculated using the open-source code MagIC (\url{https://magic-sph.github.io}). We are very grateful to anonymous reviewers for their constructive comments, which have helped improve the paper.

\backsection[Funding]{}
This study was supported by the National Key R\&D Program of China (Grant No. 2022YFF0503200), the National Natural Science Foundation of China (12250012, 42142034, 42350002), the B-type Strategic Priority Program of the CAS (XDB41000000) and the Pearl River Program (2019QN01X189).
Numerical calculations were performed on the Taiyi cluster supported by the Center for Computational Science and Engineering of Southern University of Science and Technology.
\backsection[Declaration of interests]{}
The authors report no conflict of interest.




\appendix
\section{List of numerical simulations}\label{AppA}
Table \ref{tab:data} listed all of numerical models presented in this study with detailed control parameters, diagnostic parameters and numerical resolutions. We also created an Excel spreadsheet with more comprehensive data, which is made available on Zenodo (\url{https://zenodo.org/records/12696528}).

\begin{center}
\begin{longtable}{lcccccccc}
 \multicolumn{5}{r}{{TABLE 1. (continued)}}&&&  \\ \toprule
\endfoot
\caption{Summary of the control parameters and diagnostic quantities of all numerical models in this study. The columns from left to right indicate the following: the Ekman number $E$, the modified Rayleigh number $Ra^{*}$, the Nusselt number $Nu-1$, the non-zonal Reynolds number $Re_{non}$, the zonal Reynolds number $Re_{zon}$, length scale $\ell$, the Regime corresponding to the flow state, the grid resolution and spherical harmonics of degree $l_{max}$. Zonal Reynolds number and length scale only given for the cases in GT and WR regimes.}  \\ \toprule
\endlastfoot
\toprule No. & $E$ & $Ra^*$ & $Nu-1$ & $~Re_{non}$ &$~Re_{zon}$& ~$\ell$ & Regime & $N_{r}\times L_{max}$\\ 
\endfirsthead
\toprule  
\endhead      
       &&&$\Pran=100$&&&&\\
       1 & ~~$1\times10^{-5}$ & ~~$1.85\times10^{-5}$ & 0.0157 & ~~0.713 & ~---& ~--- & ON & $120\times120$\\
       2 & ~~$1\times10^{-5}$ & ~~$3.0\times10^{-5}$ & 0.158 & ~~0.267 & ~---& ~--- & MM & $140\times140$\\
       3 & ~~$1\times10^{-5}$ & ~~$8.0\times10^{-5}$ & 1.38 & ~~1.52 & ~0.311& ~0.081 & GT & $180\times180$\\
       4 & ~~$1\times10^{-5}$ & ~~$1.0\times10^{-4}$ & 2.06 & ~~2.07 & ~0.399& ~0.074 & GT & $200\times200$\\
       5 & ~~$1\times10^{-5}$ & ~~$1.3\times10^{-4}$ & 2.90 & ~~2.72 & ~0.522& ~0.070 & GT & $220\times220$\\
       6 & ~~$1\times10^{-5}$ & ~~$2.5\times10^{-4}$ & 6.78 & ~~5.11 & ~1.04& ~0.060 & GT & $240\times240$\\
       7 & ~~$1\times10^{-5}$ & ~~$3.0\times10^{-4}$ & 8.76 & ~~6.27 & ~1.20& ~0.060 & GT & $260\times260$\\
       8 & ~~$1\times10^{-5}$ & ~~$4.0\times10^{-4}$ & 11.93 & ~~8.09 & ~1.99& ~0.059 & GT & $280\times280$\\
       9 & ~~$1\times10^{-5}$ & ~~$5.5\times10^{-4}$ & 16.02 & ~~10.53 & ~2.87& ~0.058 & GT & $300\times300$\\
       10 & ~~$1\times10^{-5}$ & ~~$1.0\times10^{-3}$ & 26.21 & ~~17.45 & ~3.87& ~0.055 & GT & $360\times360$\\
       11 & ~~$1\times10^{-5}$ & ~~$5.0\times10^{-3}$ & 58.33 & ~~53.16 & ~9.97& ~0.052 & GT & $420\times420$\\
       &&&$\Pran=7$&&&&\\
       12 & ~~$1\times10^{-4}$ & ~~$1.35\times10^{-3}$ & 0.0057 & ~~0.316 & ~---& ~--- & ON & $80\times55$\\
       13 & ~~$1\times10^{-4}$ & ~~$1.5\times10^{-3}$ & 0.0418 & ~~0.919 & ~---& ~--- & ON & $80\times55$\\
       14 & ~~$1\times10^{-4}$ & ~~$2.0\times10^{-3}$ & 0.158 & ~~2.12 & ~---& ~--- & MM & $80\times55$\\
       15 & ~~$1\times10^{-4}$ & ~~$3.0\times10^{-3}$ & 0.318 & ~~3.27 & ~---& ~--- & MM & $80\times60$\\
       16 & ~~$1\times10^{-4}$ & ~~$5.8\times10^{-3}$ & 1.04 & ~~8.85 & ~2.05& ~0.182 & GT & $100\times90$\\
       17 & ~~$1\times10^{-4}$ & ~~$8.0\times10^{-3}$ & 1.59 & ~~12.31 & ~2.85& ~0.161 & GT & $120\times90$\\
       18 & ~~$1\times10^{-4}$ & ~~$1.3\times10^{-2}$ & 2.76 & ~~19.02 & ~4.95& ~0.149 & GT & $120\times110$\\
       19 & ~~$1\times10^{-4}$ & ~~$1.7\times10^{-3}$ & 3.79 & ~~24.37 & ~6.13& ~0.145 & GT & $120\times120$\\
       20 & ~~$1\times10^{-4}$ & ~~$2.0\times10^{-3}$ & 4.68 & ~~29.04 & ~6.09& ~0.141 & GT & $120\times120$\\
       21 & ~~$1\times10^{-4}$ & ~~$3.0\times10^{-3}$ & 6.66 & ~~41.12 & ~7.20& ~0.136 & GT & $150\times150$\\
       22 & ~~$1\times10^{-4}$ & ~~$4.0\times10^{-2}$ & 8.09 & ~~51.29 & ~8.80& ~0.136 & GT & $180\times180$\\
       23 & ~~$1\times10^{-4}$ & ~~$2.0\times10^{-1}$ & 17.23 & ~~158.7 & ~35.06& ~0.150 & WR & $260\times260$\\
       24 & ~~$1\times10^{-4}$ & ~~$4.0\times10^{-1}$ & 21.49 & ~~245.7 & ~90.04& ~0.169 & WR & $320\times320$\\
       25 & ~~$3\times10^{-5}$ & ~~$5.4\times10^{-4}$ & 0.0098 & ~~0.58 & ~---& ~--- & ON & $96\times96$\\
       26 & ~~$3\times10^{-5}$ & ~~$5.5\times10^{-4}$ & 0.0147 & ~~0.72 & ~---& ~--- & ON & $96\times96$\\
       27 & ~~$3\times10^{-5}$ & ~~$6.0\times10^{-4}$ & 0.0302 & ~~1.08 & ~---& ~--- & MM & $96\times96$\\
       28 & ~~$3\times10^{-5}$ & ~~$1.5\times10^{-3}$ & 0.466 & ~~6.87 & ~---& ~--- & MM & $120\times120$\\
       29 & ~~$3\times10^{-5}$ & ~~$2.25\times10^{-3}$ & 1.01 & ~~12.87 & ~2.88& ~0.129 & GT & $140\times120$\\
       30 & ~~$3\times10^{-5}$ & ~~$3.0\times10^{-3}$ & 1.58 & ~~17.89 & ~3.96& ~0.113 & GT & $160\times160$\\
       31 & ~~$3\times10^{-5}$ & ~~$4.0\times10^{-3}$ & 2.28 & ~~23.72 & ~5.23& ~0.104 & GT & $180\times180$\\
       32 & ~~$3\times10^{-5}$ & ~~$5.0\times10^{-3}$ & 2.93 & ~~28.81 & ~6.44& ~0.099 & GT & $200\times200$\\
       33 & ~~$3\times10^{-5}$ & ~~$7.0\times10^{-3}$ & 4.95 & ~~41.93 & ~8.08& ~0.096 & GT & $260\times260$\\
       34 & ~~$3\times10^{-5}$ & ~~$1.0\times10^{-2}$ & 7.36 & ~~59.08 & ~8.96& ~0.092 & GT & $280\times280$\\
       35 & ~~$3\times10^{-5}$ & ~~$1.35\times10^{-2}$ & 9.60 & ~~76.53 & ~11.27& ~0.092 & GT & $320\times320$\\
       36 & ~~$3\times10^{-5}$ & ~~$2.0\times10^{-2}$ & ~12.86 & ~~104.6 & ~20.02& ~0.094 & GT & $360\times360$\\
       37 & ~~$3\times10^{-5}$ & ~~$5.0\times10^{-2}$ & ~21.99 & ~~211.0 & ~68.09& ~0.105 & GT & $400\times400$\\
       38 & ~~$1\times10^{-5}$ & ~~$2.4\times10^{-4}$ & ~0.011 & ~~0.823 & ~---& ~--- & ON & $120\times100$\\
       39 & ~~$1\times10^{-5}$ & ~~$2.5\times10^{-4}$ & ~0.019 & ~~1.12 & ~---& ~--- & ON & $120\times100$\\
       40 & ~~$1\times10^{-5}$ & ~~$3.5\times10^{-4}$ & ~0.100 & ~~3.11 & ~---& ~--- & MM & $160\times120$\\
       41 & ~~$1\times10^{-5}$ & ~~$5.0\times10^{-4}$ & ~0.259 & ~~5.82 & ~---& ~--- & MM & $200\times160$\\
       42 & ~~$1\times10^{-5}$ & ~~$7.0\times10^{-4}$ & ~0.504 & ~~10.16 & ~---& ~--- & MM & $200\times160$\\
       43 & ~~$1\times10^{-5}$ & ~~$1.0\times10^{-3}$ & ~1.00 & ~~18.19 & ~3.96& ~0.089 & GT & $200\times200$\\
       44 & ~~$1\times10^{-5}$ & ~~$1.3\times10^{-3}$ & ~1.60 & ~~25.43 & ~5.42& ~0.079 & GT& $200\times200$\\
       45 & ~~$1\times10^{-5}$ & ~~$2.0\times10^{-3}$ & ~2.85 & ~~39.54 & ~8.00& ~0.070 & GT & $240\times240$\\
       46 & ~~$1\times10^{-5}$ & ~~$3.0\times10^{-3}$ & ~5.39 & ~~61.77 & ~9.73& ~0.066 & GT & $320\times320$\\
       47 & ~~$1\times10^{-5}$ & ~~$4.0\times10^{-3}$ & ~7.83 & ~~83.35 & ~12.24& ~0.065 & GT & $320\times350$\\
       48 & ~~$1\times10^{-5}$ & ~~$5.0\times10^{-3}$ & ~9.96 & ~~103.3 & ~13.64& ~0.064 & GT & $400\times400$\\
       49 & ~~$1\times10^{-5}$ & ~~$1.0\times10^{-2}$ & ~18.69 & ~~196.6 & ~25.42& ~0.069 & GT & $420\times420$\\
       50 & ~~$1\times10^{-5}$ & ~~$3.0\times10^{-2}$ & ~36.95 & ~~487.8 & ~111.7& ~0.085 & GT & $480\times460$\\
       51 & ~~$3\times10^{-6}$ & ~~$9.8\times10^{-5}$ & ~0.0073 & ~~0.933 & ~---& ~--- & ON & $160\times160$\\      
       52 & ~~$3\times10^{-6}$ & ~~$7.0\times10^{-4}$ & ~2.39 & ~~50.39 & ~9.85& ~0.049 & GT & $250\times250$\\
       53 & ~~$3\times10^{-6}$ & ~~$1.0\times10^{-3}$ & ~4.36 & ~~75.46 & ~13.08& ~0.042 & GT & $320\times320$\\   
       54 & ~~$3\times10^{-6}$ & ~~$1.5\times10^{-3}$ & ~7.64 & ~~116.2 & 18.77& ~0.043 & GT & $360\times360$\\   
       55 & ~~$3\times10^{-6}$ & ~~$2.5\times10^{-3}$ & ~13.97 & ~~197.0 & ~38.23& ~0.045 & GT & $400\times400$\\   
       56 & ~~$3\times10^{-6}$ & ~~$3.5\times10^{-3}$ & ~20.34 & ~~278.2 & ~60.96& ~0.048 & GT & $480\times480$\\      
       57 & ~~$3\times10^{-6}$ & ~~$6.0\times10^{-3}$ & ~32.67 & ~~473.0 & ~189.1& ~0.058 & GT & $600\times560$\\    
       &&&$\Pran=1$&&&\\
       58 & ~~$1\times10^{-4}$ & ~~$8.0\times10^{-3}$ & ~0.036 & ~~5.396 & ~---& ~--- & ON & $80\times50$\\
       59 & ~~$1\times10^{-4}$ & ~~$1.0\times10^{-2}$ & ~0.086 & ~~9.374 & ~---& ~--- & ON & $100\times60$\\
       60 & ~~$1\times10^{-4}$ & ~~$1.4\times10^{-2}$ & ~0.158 & ~~13.90 & ~---& ~--- & ON & $120\times70$\\
       61 & ~~$1\times10^{-4}$ & ~~$2.0\times10^{-2}$ & ~0.294 & ~~23.04 & ~---& ~--- & ON & $120\times70$\\
       62 & ~~$1\times10^{-4}$ & ~~$3.0\times10^{-2}$ & ~0.515 & ~~39.10 & ~---& ~--- & MM & $120\times80$\\
       63 & ~~$1\times10^{-4}$ & ~~$4.0\times10^{-2}$ & ~0.855 & ~~57.02 & ~---& ~--- & MM & $120\times100$\\
       64 & ~~$1\times10^{-4}$ & ~~$5.0\times10^{-2}$ & ~1.18 & ~~73.31 & ~25.01& ~0.196 & GT & $120\times110$\\
       65 & ~~$1\times10^{-4}$ & ~~$6.5\times10^{-2}$ & ~1.65 & ~~97.17 & ~31.66& ~0.190 & GT & $140\times140$\\
       66 & ~~$1\times10^{-4}$ & ~~$8.0\times10^{-2}$ & ~2.28 & ~~123.9 & ~42.61& ~0.194 & GT & $160\times160$\\
       67 & ~~$1\times10^{-4}$ & ~~$1.0\times10^{-1}$ & ~3.19 & ~~159.0 & ~59.89& ~0.201 & GT & $160\times160$\\
       68 & ~~$1\times10^{-4}$ & ~~$2.0\times10^{-1}$ & ~6.08 & ~~282.9 & ~138.4& ~0.212 & WR & $200\times200$\\
       69 & ~~$1\times10^{-4}$ & ~~$3.0\times10^{-1}$ & ~7.76 & ~~373.2 & ~213.9& ~0.223 & WR & $240\times240$\\
       70 & ~~$3\times10^{-5}$ & ~~$2.6\times10^{-3}$ & ~0.0035 & ~~2.078 & ~---& ~--- & ON & $100\times63$\\
       71 & ~~$3\times10^{-5}$ & ~~$3.0\times10^{-3}$ & ~0.0310 & ~~6.806 & ~---& ~--- & ON & $120\times90$\\
       72 & ~~$3\times10^{-5}$ & ~~$4.5\times10^{-3}$ & ~0.104 & ~~14.31 & ~---& ~--- & ON & $120\times100$\\
       73 & ~~$3\times10^{-5}$ & ~~$4.8\times10^{-3}$ & ~0.116 & ~~15.49 & ~---& ~--- & MM & $120\times100$\\
       74 & ~~$3\times10^{-5}$ & ~~$8.0\times10^{-3}$ & ~0.296 & ~~34.20 & ~---& ~--- & MM & $140\times110$\\
       75 & ~~$3\times10^{-5}$ & ~~$1.5\times10^{-2}$ & ~0.788 & ~~81.48 & ~---& ~--- & MM & $180\times150$\\
       76 & ~~$3\times10^{-5}$ & ~~$2.0\times10^{-2}$ & ~1.22 & ~~114.2 & ~37.00& ~0.143 & GT & $200\times180$\\
       77 & ~~$3\times10^{-5}$ & ~~$2.5\times10^{-2}$ & ~1.68 & ~~146.6 & ~45.79& ~0.136 & GT & $200\times200$\\
       78 & ~~$3\times10^{-5}$ & ~~$3.5\times10^{-2}$ &~ 2.74 & ~~219.5 & ~74.01& ~0.145 & GT & $240\times200$\\
       79 & ~~$3\times10^{-5}$ & ~~$5.0\times10^{-2}$ & ~4.86 & ~~334.5 & ~136.2& ~0.156 & GT & $240\times220$\\
       80 & ~~$3\times10^{-5}$ & ~~$7.0\times10^{-2}$ & ~7.15 & ~~455.1 & ~227.1& ~0.165 & GT & $280\times260$\\
       81 & ~~$3\times10^{-5}$ & ~~$8.0\times10^{-2}$ & ~8.071 & ~~515.4 & ~255.8& ~0.168 & WR & $280\times260$\\
       82 & ~~$3\times10^{-5}$ & ~~$1.0\times10^{-1}$ & ~9.578 & ~~602.5 & ~355.3& ~0.176 & WR & $320\times320$\\
       83 & ~~$1\times10^{-5}$ & ~~$1.09\times10^{-3}$ & ~0.0036 & ~~2.82 & ~---& ~--- & ON & $120\times100$\\
       84 & ~~$1\times10^{-5}$ & ~~$1.1\times10^{-3}$ & ~0.0058 & ~~3.58 & ~---& ~--- & ON & $160\times100$\\
       85 & ~~$1\times10^{-5}$ & ~~$1.2\times10^{-3}$ & ~0.019 & ~~6.88 & ~---& ~--- & ON & $160\times100$\\
       86 & ~~$1\times10^{-5}$ & ~~$1.38\times10^{-3}$ & ~0.038 & ~~10.28 & ~---& ~--- & ON & $160\times100$\\
       87 & ~~$1\times10^{-5}$ & ~~$1.5\times10^{-3}$ & ~0.049 & ~~11.70 & ~---& ~--- & ON & $160\times100$\\
       88 & ~~$1\times10^{-5}$ & ~~$1.7\times10^{-3}$ & ~0.069 & ~~14.59 & ~---& ~--- & ON & $160\times130$\\
       89 & ~~$1\times10^{-5}$ & ~~$1.73\times10^{-3}$ & ~0.074 & ~~15.45 & ~---& ~--- & MM & $180\times130$\\    
       90 & ~~$1\times10^{-5}$ & ~~$3.0\times10^{-3}$ & ~0.242 & ~~40.82 & ~---& ~--- & MM & $200\times150$\\
       91 & ~~$1\times10^{-5}$ & ~~$5.5\times10^{-3}$ & ~0.610 & ~~96.06 & ~---& ~--- & MM & $280\times180$\\
       92 & ~~$1\times10^{-5}$ & ~~$8.0\times10^{-3}$ & ~1.02 & ~~148.6 & ~47.05& ~0.110 & GT & $280\times190$\\
       93 & ~~$1\times10^{-5}$ & ~~$1.2\times10^{-2}$ & ~1.86 & ~~239.5 & ~66.55& ~0.103 & GT & $320\times240$\\
       94 & ~~$1\times10^{-5}$ & ~~$1.5\times10^{-2}$ & ~2.64 & ~~318.0 & ~86.78& ~0.105 & GT & $320\times240$\\
       95 & ~~$1\times10^{-5}$ & ~~$2.0\times10^{-2}$ & ~4.49 & ~~470.0 & ~187.7& ~0.122 & GT & $320\times300$\\
       96 & ~~$1\times10^{-5}$ & ~~$2.5\times10^{-2}$ & ~6.30 & ~~609.1 & ~257.0& ~0.124 & GT & $320\times300$\\
       97 & ~~$1\times10^{-5}$ & ~~$3.0\times10^{-2}$ & ~8.08 & ~~738.9 & ~342.2& ~0.130 & GT & $320\times300$\\
       98 & ~~$1\times10^{-5}$ & ~~$3.5\times10^{-2}$ & ~9.74 & ~~862.0 & ~415.0& ~0.132 & GT & $320\times320$\\
       99 & ~~$1\times10^{-5}$ & ~~$5.0\times10^{-2}$ & ~13.30 & ~~1126 & ~757.0& ~0.148 & GT & $320\times320$\\
       100 & ~~$3\times10^{-6}$ & ~~$4.3\times10^{-4}$ & ~0.00344 & ~~3.80 & ~---& ~--- & ON & $160\times100$\\
       101 & ~~$3\times10^{-6}$ & ~~$5.0\times10^{-4}$ & ~0.0207 & ~~9.88 & ~---& ~--- & ON & $160\times100$\\
       102 & ~~$3\times10^{-6}$ & ~~$6.0\times10^{-4}$ & ~0.0434 & ~~15.2 & ~---& ~--- & MM & $180\times120$\\
       103 & ~~$3\times10^{-6}$ & ~~$1.0\times10^{-3}$ & ~0.184 & ~~45.5 & ~---& ~--- & MM & $200\times200$\\
       104 & ~~$3\times10^{-6}$ & ~~$4.5\times10^{-3}$ & ~1.54 & ~~317.0 & ~96.96& ~0.074 & GT & $260\times260$\\
       105 & ~~$3\times10^{-6}$ & ~~$6.0\times10^{-3}$ & ~2.46 & ~~460.8 & ~138.7& ~0.076 & GT & $300\times300$\\
       106 & ~~$3\times10^{-6}$ & ~~$8.0\times10^{-3}$ & ~4.22 & ~~716.4 & ~217.5& ~0.086 & GT & $320\times320$\\
       107 & ~~$3\times10^{-6}$ & ~~$1.2\times10^{-2}$ & ~8.32 & ~~1203 & ~544.2& ~0.102 & GT & $360\times360$\\
       108 & ~~$3\times10^{-6}$ & ~~$1.5\times10^{-2}$ & ~11.20 & ~~1521 & ~761.9& ~0.108 & GT & $400\times400$\\
       109 & ~~$3\times10^{-6}$ & ~~$3.6\times10^{-2}$ & ~24.64 & ~~2551 & ~3320& ~0.157 & GT & $480\times480$\\
       &&&$\Pran=0.1$&&&\\
       110 & ~~$1\times10^{-4}$ & ~~$2.95\times10^{-2}$ & ~0.0027 & ~~10.56 & ~---& ~--- & ON & $100\times50$\\
       111 & ~~$1\times10^{-4}$ & ~~$3.0\times10^{-2}$ & ~0.0039 & ~~12.73 & ~---& ~--- & ON & $100\times50$\\
       112 & ~~$1\times10^{-4}$ & ~~$3.2\times10^{-2}$ & ~0.0077 & ~~18.35 & ~---& ~--- & ON & $100\times55$\\
       113& ~~$1\times10^{-4}$ & ~~$3.8\times10^{-2}$ & ~0.016 & ~~28.44 & ~---& ~--- & ON & $100\times70$\\
       114 & ~~$1\times10^{-4}$ & ~~$4.0\times10^{-2}$ & ~0.017 & ~~30.16 & ~---& ~--- & ON & $100\times70$\\
       115 & ~~$1\times10^{-4}$ & ~~$4.6\times10^{-2}$ & ~0.020 & ~~35.47 & ~---& ~--- & MM & $100\times70$\\
       116 & ~~$1\times10^{-4}$ & ~~$6.0\times10^{-2}$ & ~0.044 & ~~57.47 & ~42.88& ~0.517 & GT & $100\times80$\\
       117 & ~~$1\times10^{-4}$ & ~~$7.0\times10^{-2}$ & ~0.059 & ~~69.24 & ~56.57& ~0.530 & GT & $120\times80$\\
       118 & ~~$1\times10^{-4}$ & ~~$9.0\times10^{-2}$ & ~0.097 & ~~97.27 & ~82.49& ~0.555 & GT & $120\times100$\\
       119 & ~~$1\times10^{-4}$ & ~~$1.0\times10^{-1}$ & ~0.106 & ~~104.5 & ~94.99& ~0.557 & GT & $150\times100$\\
       120 & ~~$1\times10^{-4}$ & ~~$2.0\times10^{-1}$ & ~0.487 & ~~282.2 & ~258.2& ~0.420 & GT & $200\times140$\\
       121 & ~~$1\times10^{-4}$ & ~~$3.0\times10^{-1}$ & ~1.053 & ~~494.9 & ~435.8& ~0.375 & WR & $250\times180$\\
       122 & ~~$1\times10^{-4}$ & ~~$4.35\times10^{-1}$ & ~1.734 & ~~750.9 & ~575.8& ~0.338 & WR & $280\times200$\\
       123 & ~~$3\times10^{-5}$ & ~~$9.5\times10^{-3}$ & ~0.0011 & ~~8.96 & ~---& ~--- & ON & $100\times70$\\
       124 & ~~$3\times10^{-5}$ & ~~$9.8\times10^{-2}$ & ~0.0028 & ~~14.29 & ~---& ~--- & ON & $100\times70$\\
       125 & ~~$3\times10^{-5}$ & ~~$1.0\times10^{-2}$ & ~0.0038 & ~~16.69& ~---& ~--- & ON & $100\times80$\\
       126 & ~~$3\times10^{-5}$ & ~~$1.1\times10^{-2}$ & ~0.0074 & ~~23.83 & ~---& ~--- & ON & $100\times80$\\
       127 & ~~$3\times10^{-5}$ & ~~$1.2\times10^{-2}$ & ~0.0095 & ~~27.58 & ~---& ~--- & ON & $120\times80$\\
       128 & ~~$3\times10^{-5}$ & ~~$1.4\times10^{-2}$ & ~0.0130 & ~~34.76 & ~---& ~--- & MM & $120\times100$\\
       129 & ~~$3\times10^{-5}$ & ~~$1.8\times10^{-2}$ & ~0.0393 & ~~67.03 & ~50.93& ~0.350 & GT & $120\times120$\\
       130 & ~~$3\times10^{-5}$ & ~~$2.0\times10^{-2}$ & ~0.049 & ~~79.30 & ~60.20& ~0.364 & GT & $140\times120$\\
       131 & ~~$3\times10^{-5}$ & ~~$2.5\times10^{-2}$ & ~0.076 & ~~105.6 & ~91.83& ~0.375 & GT & $140\times130$\\
       132 & ~~$3\times10^{-5}$ & ~~$3.0\times10^{-2}$ & ~0.107 & ~~134.2 & ~126.2& ~0.391 & GT & $140\times130$\\
       133 & ~~$3\times10^{-5}$ & ~~$5.5\times10^{-2}$ & ~0.287 & ~~285.2 & ~301.9& ~0.422 & GT & $160\times140$\\
       134 & ~~$3\times10^{-5}$ & ~~$8.0\times10^{-2}$ & ~0.593 & ~~480.3 & ~505.7& ~0.376 & GT & $180\times150$\\
       135 & ~~$3\times10^{-5}$ & ~~$1.1\times10^{-1}$ & ~1.10 & ~~738.0 & ~769.7& ~0.335 & GT & $200\times180$\\
       136 & ~~$3\times10^{-5}$ & ~~$2.0\times10^{-1}$ & ~2.43 & ~~1380 & ~1391& ~0.297 & WR & $240\times200$\\
       137 & ~~$3\times10^{-5}$ & ~~$3.0\times10^{-1}$ & ~3.52 & ~~1846 & ~1876& ~0.275 & WR & $280\times250$\\      
       138 & ~~$1\times10^{-5}$ & ~~$3.6\times10^{-3}$ & ~0.00099 & ~~11.10 & ~---& ~--- & ON & $160\times100$\\
       139 & ~~$1\times10^{-5}$ & ~~$3.7\times10^{-3}$ & ~0.0022 & ~~16.51 & ~---& ~--- & ON & $160\times100$\\
       140 & ~~$1\times10^{-5}$ & ~~$4.0\times10^{-3}$ & ~0.0050 & ~~25.20 & ~---& ~--- & ON & $160\times100$\\
       141 & ~~$1\times10^{-5}$ & ~~$4.44\times10^{-3}$ & ~0.0074 & ~~30.97 & ~---& ~--- & ON &$200\times130$\\
       142 & ~~$1\times10^{-5}$ & ~~$5.1\times10^{-3}$ & ~0.010 & ~~37.05 & ~---& ~--- & MM & $200\times130$\\
       143 & ~~$1\times10^{-5}$ & ~~$7.1\times10^{-3}$ & ~0.053 & ~~107.0 & ~75.23& ~0.244 & GT & $224\times170$\\
       144 & ~~$1\times10^{-5}$ & ~~$1.0\times10^{-2}$ & ~0.115 & ~~182.0 & ~155.5& ~0.272 & GT & $224\times180$\\
       145 & ~~$1\times10^{-5}$ & ~~$1.3\times10^{-2}$ & ~0.187 & ~~266.0 & ~245.0& ~0.293 & GT & $240\times180$\\
       146 & ~~$1\times10^{-5}$ & ~~$2.0\times10^{-2}$ & ~0.332 & ~~441.3 & ~445.9& ~0.307 & GT & $280\times190$\\
       147 & ~~$1\times10^{-5}$ & ~~$3.0\times10^{-2}$ & ~0.558 & ~~688.4 & ~727.5& ~0.328 & GT & $280\times200$\\
       148 & ~~$1\times10^{-5}$ & ~~$4.0\times10^{-2}$ & ~0.920 & ~~973.1 & ~1044& ~0.299 & GT & $280\times200$\\
       149 & ~~$1\times10^{-5}$ & ~~$5.0\times10^{-2}$ & ~1.39 & ~~1227 & ~1421& ~0.279 & GT & $320\times200$\\
       150 & ~~$1\times10^{-5}$ & ~~$6.5\times10^{-2}$ & ~1.98 & ~~1615 & ~1957& ~0.277 & ~GT & $320\times200$\\      
       151 & ~~$1\times10^{-5}$ & ~~$1.0\times10^{-1}$ & ~3.40 & ~~2508 & ~2848& ~0.262 & WR & $320\times250$\\
       152 & ~~$1\times10^{-5}$ & ~~$3.0\times10^{-1}$ & ~8.21 & ~~5087 & ~6366& ~0.226 & WR & $360\times260$\\
       153 & ~~$3\times10^{-6}$ & ~~$1.3\times10^{-3}$ & ~0.0012 & ~~16.61 & ~---& ~--- & ON & $180\times100$\\
       154 & ~~$3\times10^{-6}$ & ~~$1.5\times10^{-3}$ & ~0.0052 & ~~33.66 & ~---& ~--- & ON & $180\times100$\\
       155 & ~~$3\times10^{-6}$ & ~~$1.7\times10^{-3}$ & ~0.0070 & ~~38.68 & ~---& ~--- & ON & $180\times120$\\
       156 & ~~$3\times10^{-6}$ & ~~$3.0\times10^{-3}$ & ~0.104 & ~~231.8 & ~172.6& ~0.180 & GT & $200\times180$\\
       157 & ~~$3\times10^{-6}$ & ~~$6.0\times10^{-3}$ & ~0.31 & ~~617.6 & ~515.4& ~0.221 & GT & $220\times200$\\
       158 & ~~$3\times10^{-6}$ & ~~$1.0\times10^{-2}$ & ~0.52 & ~~1030 & ~927.0& ~0.243 & GT & $280\times250$\\
       159 & ~~$3\times10^{-6}$ & ~~$1.5\times10^{-2}$ & ~0.88 & ~~1558 & ~1494& ~0.235 & GT & $320\times280$\\
       160 & ~~$3\times10^{-6}$ & ~~$2.0\times10^{-2}$ & ~1.44 & ~~2148 & ~2159& ~0.221 & GT & $320\times320$\\
       161 & ~~$3\times10^{-6}$ & ~~$3.0\times10^{-2}$ & ~2.65 & ~~3171 & ~3707& ~0.221 & GT & $320\times320$\\
       162 & ~~$3\times10^{-6}$ & ~~$4.0\times10^{-2}$ & ~3.92 & ~~4072 & ~5176& ~0.218 & GT & $380\times380$\\
       163 & ~~$3\times10^{-6}$ & ~~$5.0\times10^{-2}$ & ~5.05 & ~~5215 & ~5827& ~0.200 & GT & $400\times400$\\
       164 & ~~$1\times10^{-6}$ & ~~$1.0\times10^{-2}$ & ~1.478 & ~~3191 & ~1868& ~0.141 & GT & $380\times380$\\
       165 & ~~$1\times10^{-6}$ & ~~$2.5\times10^{-2}$ & ~6.521 & ~~9179 & ~10914& ~0.164 & GT & $500\times450$\\
       &&&$\Pran=0.01$&&&\\
       166 & ~~$1\times10^{-4}$ & ~~$1.79\times10^{-1}$ & ~0.00032 & ~~30.75 & ~---& ~--- & ON & $100\times60$\\
       167 & ~~$1\times10^{-4}$ & ~~$1.83\times10^{-1}$ & ~0.00044 & ~~36.63 & ~---& ~--- & ON & $100\times60$\\
       168 & ~~$1\times10^{-4}$ & ~~$1.9\times10^{-1}$ & ~0.00062 & ~~43.60 & ~---& ~--- & ON & $100\times80$\\
       169 & ~~$1\times10^{-4}$ & ~~$2.0\times10^{-1}$ & ~0.00082 & ~~50.93 & ~---& ~--- & ON & $100\times80$\\
       170 & ~~$1\times10^{-4}$ & ~~$2.35\times10^{-1}$ & ~0.0020 & ~~80.47 & ~---& ~--- & ON & $100\times80$\\
       171 & ~~$1\times10^{-4}$ & ~~$2.5\times10^{-1}$ & ~0.0025 & ~~94.74 & ~---& ~--- & MM & $100\times80$\\
       172 & ~~$1\times10^{-4}$ & ~~$2.8\times10^{-1}$ & ~0.028 & ~~307.6 & ~299.8& ~0.680 & GT & $100\times100$\\
       173 & ~~$1\times10^{-4}$ & ~~$3.2\times10^{-1}$ & ~0.044 & ~~400.8 & ~395.0& ~0.680 & GT & $100\times100$\\
       174 & ~~$1\times10^{-4}$ & ~~$3.6\times10^{-1}$ & ~0.066 & ~~513.1 & ~503.8& ~0.665 & WR & $120\times120$\\
       175 & ~~$1\times10^{-4}$ & ~~$6.0\times10^{-1}$ & ~0.181 & ~~1045 & ~920.4& ~0.566 & WR & $160\times160$\\
       176 & ~~$1\times10^{-4}$ & ~~$1.0\times10^{0}$ & ~0.058 & ~~2228 & ~1245& ~0.396 & WR & $240\times240$\\
       177 & ~~$3\times10^{-5}$ & ~~$5.5\times10^{-2}$ & ~0.00021 & ~~33.3 & ~---& ~--- & ON & $120\times90$\\
       178 & ~~$3\times10^{-5}$ & ~~$6.0\times10^{-2}$ & ~0.00072 & ~~62.6 & ~---& ~--- & ON & $120\times100$\\
       179 & ~~$3\times10^{-5}$ & ~~$7.0\times10^{-2}$ & ~0.0017 & ~~98.0 & ~---& ~--- & ON &~$120\times110$\\
       180 & ~~$3\times10^{-5}$ & ~~$8.0\times10^{-2}$ & ~0.0022 & ~~115.1 & ~---& ~--- & ON & $120\times120$\\
       181 & ~~$3\times10^{-5}$ & ~~$8.4\times10^{-2}$ & ~0.0031 & ~~134.6 & ~---& ~--- & MM & $120\times120$\\
       182 & ~~$3\times10^{-5}$ & ~~$8.5\times10^{-2}$ & ~0.025 & ~~379.5 & ~388.1& ~0.600 & GT & $140\times127$\\
       183 & ~~$3\times10^{-5}$ & ~~$1.2\times10^{-1}$ & ~0.049 & ~~568.7 & ~714.8& ~0.635 & GT & $160\times150$\\
       184 & ~~$3\times10^{-5}$ & ~~$2.0\times10^{-1}$ & ~0.138 & ~~1157 & ~1424& ~0.560 & GT & $200\times200$\\
       185 & ~~$3\times10^{-5}$ & ~~$3.0\times10^{-1}$ & ~0.296 & ~~1928 & ~2194& ~0.452 & WR & $240\times240$\\
       186 & ~~$1\times10^{-5}$ & ~~$2.0\times10^{-2}$ & ~0.00018 & ~~40.67 & ~---& ~--- & ON & $160\times90$\\
       187 & ~~$1\times10^{-5}$ & ~~$2.1\times10^{-2}$ & ~0.00049 & ~~68.87 & ~---& ~--- & ON & $160\times90$\\
       188 & ~~$1\times10^{-5}$ & ~~$2.2\times10^{-2}$ & ~0.00072 & ~~84.81 & ~---& ~--- & ON & $160\times120$\\
       189 & ~~$1\times10^{-5}$ & ~~$2.4\times10^{-2}$ & ~0.0014 & ~~118.0 & ~---& ~--- & ON & $160\times120$\\
       190 & ~~$1\times10^{-5}$ & ~~$2.6\times10^{-2}$ & ~0.0020 & ~~142.0 & ~---& ~--- & ON & $160\times120$\\
       191 & ~~$1\times10^{-5}$ & ~~$2.7\times10^{-2}$ & ~0.0025 & ~~157.0 & ~---& ~--- & MM & $180\times130$\\
       192 & ~~$1\times10^{-5}$ & ~~$2.83\times10^{-2}$ & ~0.0215 & ~~442.7 & ~479.4& ~0.490 & GT & $200\times130$\\
       193 & ~~$1\times10^{-5}$ & ~~$4.0\times10^{-2}$ & ~0.048 & ~~693.7 & ~940.4& ~0.527 & GT & $200\times130$\\
       194 & ~~$1\times10^{-5}$ & ~~$5.8\times10^{-2}$ & ~0.101 & ~~1183 & ~1634& ~0.529 & GT & $320\times200$\\
       195 & ~~$1\times10^{-5}$ & ~~$7.5\times10^{-2}$ & ~0.150 & ~~1612 & ~2247& ~0.518 & GT & $320\times230$\\
       196 & ~~$1\times10^{-5}$ & ~~$1.0\times10^{-1}$ & ~0.232 & ~~2240 & ~3157& ~0.488 & GT & $320\times230$\\
       197 & ~~$1\times10^{-5}$ & ~~$2.0\times10^{-1}$ & ~0.651 & ~~4475 & ~5973& ~0.355 & WR & $320\times240$\\
       198 & ~~$1\times10^{-5}$ & ~~$3.0\times10^{-1}$ & ~1.15 & ~~6396 & ~8146& ~0.296 & WR & $360\times300$\\
       199 & ~~$1\times10^{-5}$ & ~~$5.0\times10^{-1}$ & ~2.13 & ~~9782 & ~9908& ~0.212 & WR & $400\times360$\\
       200 & ~~$3\times10^{-6}$ & ~~$7.0\times10^{-3}$ & ~0.00025 & ~~68.72 & ~---& ~--- & ON & $160\times100$\\
       201 & ~~$3\times10^{-6}$ & ~~$7.2\times10^{-3}$ & ~0.00055 & ~~104.6 & ~---& ~--- & ON & $160\times100$\\
       202 & ~~$3\times10^{-6}$ & ~~$8.0\times10^{-3}$ & ~0.00095 & ~~131.2 & ~---& ~--- & MM & $160\times120$\\
       203 & ~~$3\times10^{-6}$ & ~~$9.2\times10^{-3}$ & ~0.0189 & ~~564.7 & ~587.4& ~0.358 & GT & $200\times160$\\
       204 & ~~$3\times10^{-6}$ & ~~$1.0\times10^{-2}$ & ~0.035 & ~~757.7 & ~566.9& ~0.365 & GT & $240\times180$\\
       205 & ~~$3\times10^{-6}$ & ~~$2.0\times10^{-2}$ & ~0.153 & ~~2078 & ~2825& ~0.417 & GT & $280\times260$\\      
       206 & ~~$3\times10^{-6}$ & ~~$3.0\times10^{-2}$ & ~0.239 & ~~3088 & ~4592& ~0.434 & GT & $320\times300$\\
       207 & ~~$3\times10^{-6}$ & ~~$5.0\times10^{-2}$ & ~0.427 & ~~4782 & ~7837& ~0.423 & GT & $320\times360$\\
       208 & ~~$3\times10^{-6}$ & ~~$7.0\times10^{-2}$ & ~0.678 & ~~6555 & ~11101& ~0.385 & WR & $380\times380$\\
       209 & ~~$3\times10^{-6}$ & ~~$1.0\times10^{-1}$ & ~1.04 & ~~7444 & ~14943& ~0.301 & WR & $504\times480$\\
       210 & ~~$1\times10^{-6}$ & ~~$2.5\times10^{-2}$ & ~0.640 & ~~8930 & ~15717& ~0.332 & GT & $500\times400$\\
       211 & ~~$1\times10^{-6}$ & ~~$4.5\times10^{-2}$ & ~1.19 & ~~12286 & ~28600& ~0.329 & GT & $640\times580$\\
  \label{tab:data}
\end{longtable}
\end{center}

\section{Critical Rayleigh numbers and wavenumbers }\label{AppB}
Table \ref{tab:critical} listed the critical Rayleigh numbers and wavenumbers of the onset of convection used in this study. Critical values are calculated using the the linear onset package of the open-source code MagIC \citep{wicht2002inner}, which is freely available at \url{https://magic-sph.github.io}. We have verified our calculations with previous work. For example, \cite{Barik2023} obtained the values $Ra_c=1.05567 \times 10^7$ and $m_c=15$ for $E=1 \times 10^{-5}$ and $\Pran =1$, which is in excellent with our results under the same $E$ and $\Pran$.
\begin{center}
\begin{longtable}{lcccc}
\caption{Critical modified Rayleigh numbers $Ra_{c}^{*}$, critical Rayleigh numbers $Ra_c$ and critical azimuthal wavenumbers $m_{c}$ for different Ekman $E$ and Prandtl numbers $\Pran$ employed here.}  \\ \toprule
\endlastfoot
\toprule $\Pran$ & $E$ & $Ra_{c}^{*}$ & $Ra_c$ & ~$m_{c}$
\endfirsthead     
       0.01 & ~$1\times10^{-4}$ &  ~~$1.695\times10^{-1}$& ~~$1.695\times10^5$ & ~4 \\
       0.01 & ~$3\times10^{-5}$ &  ~~$5.368\times10^{-2}$& ~~$5.964\times10^5$ & ~5 \\
       0.01 & ~$1\times10^{-5}$ & ~~ $1.955\times10^{-2}$& ~~$1.955\times10^6$ & ~6 \\
       0.01 & ~$3\times10^{-6}$ &  ~~$6.759\times10^{-3}$& ~~$7.510\times10^6$ & ~8 \\
       0.01 & ~$1\times10^{-6}$ & ~~ $2.594\times10^{-3}$& ~~$2.594\times10^7$ & ~12 \\
       ~0.1 & ~$1\times10^{-4}$ &  ~~$2.856\times10^{-2}$& ~~$2.856\times10^5$ & ~6 \\
       ~0.1 & ~$3\times10^{-5}$ &  ~~$9.351\times10^{-3}$& ~~$1.039\times10^6$ & ~8 \\
       ~0.1 & ~$1\times10^{-5}$ &  ~~$3.529\times10^{-3}$& ~~$3.529\times10^6$ & ~11 \\
       ~0.1 & ~$3\times10^{-6}$ &  ~~$1.261\times10^{-3}$& ~~$1.401\times10^7$ & ~16 \\
       ~0.1 & ~$1\times10^{-6}$ &  ~~$5.102\times10^{-4}$& ~~$5.102\times10^7$ & ~23 \\
       ~1 & ~$1\times10^{-4}$ &  ~~$6.960\times10^{-3}$& ~~$6.960\times10^5$ & 8 \\
       ~1 & ~$3\times10^{-5}$ &  ~~$2.546\times10^{-3}$& ~~$2.829\times10^6$ & ~11 \\
       ~1 & ~$1\times10^{-5}$ &  ~~$1.056\times10^{-3}$& ~~$1.056\times10^7$ & ~15 \\
       ~1 & ~$3\times10^{-6}$ &  ~~$4.145\times10^{-4}$& ~~$4.606\times10^7$ & ~22 \\
       ~7 & ~$1\times10^{-4}$ &  ~~$1.326\times10^{-3}$& ~~$9.282\times10^5$ & ~8 \\
       ~7 & ~$3\times10^{-5}$ &  ~~$5.183\times10^{-4}$& ~~$4.031\times10^6$ & ~11 \\
       ~7 & ~$1\times10^{-5}$ &  ~~$2.263\times10^{-4}$& ~~$1.584\times10^7$ & ~17 \\
       ~7 & ~$3\times10^{-6}$ &  ~~$9.302\times10^{-5}$& ~~$7.235\times10^7$ & ~25 \\
       100 & ~$1\times10^{-5}$ &  ~~$1.731\times10^{-5}$& ~~$1.731\times10^7$ & ~17 \\
\label{tab:critical}
\end{longtable}
\end{center}


\bibliographystyle{jfm}
\bibliography{jfm.bib}





\end{document}